\documentclass[twocolumn]{aastex62}

\usepackage[utf8]{inputenc}
\DeclareUnicodeCharacter{2212}{-}
\usepackage{graphicx}
\graphicspath{{./}{Figures}}
\usepackage{amsmath}
\usepackage{ulem}
\usepackage{breqn}
\usepackage{afterpage}
\usepackage{enumitem}
\usepackage{xcolor}
\usepackage{float}
\usepackage{xspace} 
\usepackage{url}
\usepackage{fontawesome}
\setenumerate{itemsep=0mm}
\usepackage{lineno}
\usepackage{wrapfig}
\usepackage{placeins}
\usepackage{CJK}
\usepackage[hang,flushmargin]{footmisc}

\newcommand{\vv}{{\tablenotemark{\footnotesize{a}}}}

\newcommand{\msun}{\ensuremath{M_\odot}\xspace}

\newcommand{\teff}{\ensuremath{{T_{\rm eff}}}\xspace}
\newcommand{\logg}{\ensuremath{{\log g}}\xspace}
\newcommand{\vlos}{\ensuremath{{v_{\rm los}}}\xspace}
\newcommand{\vturb}{\ensuremath{{\xi}}\xspace}
\newcommand{\dhelio}{\ensuremath{{d_{\rm helio}}}\xspace}
\newcommand{\cfe}{\ensuremath{{\rm[C/Fe]}}\xspace}
\newcommand{\ac}{\ensuremath{{A({\rm C})}}\xspace}
\newcommand{\sn}{\ensuremath{{S/N}}\xspace}

\newcommand{\etot}{\ensuremath{{E_{\rm tot}}}\xspace}

\newcommand{\gxp}{\textit{Gaia} XP\xspace}
\newcommand{\gdr}{GDR3\_526285\xspace} 
\newcommand{\caffaustar}{SDSS J102915$+$172927\xspace} 
\newcommand{\kellerstar}{SMSS J031300.36$-$670839.3\xspace} 

\usepackage{amsmath}
\usepackage{amssymb}
\usepackage{xspace}
\usepackage{xifthen}
\usepackage{eso-pic}

\definecolor{forestgreen}{HTML}{228B22}
\definecolor{urlblue}{HTML}{000000}

\mathchardef\mhyphen="2D

\newlength{\dhatheight}

\newcommand{\unit}[1]{\ensuremath{\mathrm{\,#1}}\xspace}

\newcommand{\km}{\unit{km}}
\newcommand{\kms}{\km \second^{-1}}

\newcommand{\second}{\unit{s}}

\newcommand{\bandvar}[2][]{%
  \ifthenelse{\isempty{#1}}{\var{#2}}{\var{#2\_#1}}%
}

\newcommand{\feh}{{\ensuremath{\rm [Fe/H]}}\xspace}

\newcommand{\var}[1]{\ensuremath{\texttt{\MakeUppercase{#1}}}\xspace}

\providecommand\physrep{\ref@jnl{Phys.~Rep.}}%
\providecommand\apjs{\ref@jnl{ApJS}}%
\providecommand{\jcap}{\ref@jnl{JCAP}}%

\shorttitle{An UMP star from \textit{Gaia} XP}
\shortauthors{Limberg et al.}

\begin{document}

\title{\textbf{
Discovery of an $\feh \sim -4.8$ Star
in \textit{Gaia} XP Spectra\footnote{Based on observations gathered with the 6.5 m Magellan Telescopes located at Las Campanas Observatory, Chile.}
}}

%

\newcommand{\MITPhysics}{Department of Physics, Massachusetts Institute of Technology, 77 Massachusetts Avenue, Cambridge, MA 02139, USA}
\newcommand{\MITKavli}{Kavli Institute for Astrophysics and Space Research, Massachusetts Institute of Technology, 77 Massachusetts Avenue, Cambridge, MA 02139, USA}
\newcommand{\UChicagoAA}{Department of Astronomy \& Astrophysics, University of Chicago, 5640 S. Ellis Avenue, Chicago, IL 60637, USA}
\newcommand{\KICP}{Kavli Institute for Cosmological Physics, University of Chicago, 5640 S Ellis Avenue, Chicago, IL 60637, USA}
\newcommand{\IAGUSP}{Universidade de S\~ao Paulo, Instituto de Astronomia, Geof\'isica e Ci\^encias Atmosf\'ericas, Departamento de Astronomia, SP 05508-090, S\~ao Paulo, Brazil}
\newcommand{\NOIRLab}{NSF NOIRLab, Tucson, AZ 85719, USA}
\newcommand{\JINA}{Joint Institute for Nuclear Astrophysics--Center for the Evolution of the Elements (JINA-CEE), USA}


\correspondingauthor{Guilherme Limberg}
\email{limberg@uchicago.edu}
\author[0000-0002-9269-8287]{Guilherme~Limberg}
\affiliation{\KICP}

\author[0000-0003-4479-1265]{Vinicius M.\ Placco}
\affiliation{\NOIRLab}

\author[0000-0002-4863-8842]{Alexander~P.~Ji}
\affiliation{\UChicagoAA}
\affiliation{\KICP}

\author{Yupeng Yao}
\affiliation{Department of Computer Science and Engineering, University of North Texas, 3940 N. Elm Street, Denton, TX 76207, USA}

\author[0000-0002-7155-679X]{Anirudh Chiti}
\affiliation{\UChicagoAA}
\affiliation{\KICP}

\author[0000-0001-9178-3992]{Mohammad K.\ Mardini}
\affiliation{Department of Physics, Zarqa University, Zarqa 13110, Jordan}
\affiliation{Jordanian Astronomical Virtual Observatory, Zarqa University, Zarqa 13110, Jordan}
\affiliation{\JINA}

\author[0000-0002-2139-7145]{Anna Frebel}
\affiliation{\MITPhysics}
\affiliation{\MITKavli}
\affiliation{\JINA}

\author[0000-0001-7479-5756]{Silvia Rossi}
\affiliation{\IAGUSP}

\begin{abstract}
We report on the discovery of \gdr (\textit{Gaia} DR3 Source ID 5262850721755411072), a star with $\feh = -4.82 \pm 0.25$ and one of the lowest metal ($\text{atomic number} > 2$) mass fractions ever found ($Z_{\rm GDR3\_526585} \lesssim 1.0 \times 10^{-6}$). We first
identified it as an ultra metal-poor (UMP; $\feh < -4$) red giant-branch (RGB) star candidate in the \textit{Gaia} BP/RP (XP) spectro-photometric catalog (\textit{Gaia} $G$ magnitude $\approx$15). A combination of multi-band photometry and high-resolution spectroscopic analysis under local thermodynamic equilibrium confirmed the status of \gdr as a distant ($\approx$24\,kpc from the Sun) RGB star ($\teff = 4596\,{\rm K}$, $\logg = 0.88$) in the Milky Way's outer halo. We obtain only an upper limit for the carbon abundance of $\rm[C/H] < -4.32$, resulting in $\rm[C/Fe] < +0.50$. A correction for the evolutionary carbon depletion ($\Delta \cfe = +0.68$) brings the nominal carbon-to-iron ratio upper limit to $\cfe_{\rm cor} < +1.18$. Given its extraordinarily low [C/H], \gdr likely formed from gas cooled via dust grains rather than fine structure line cooling. The kinematics of \gdr suggests that this star was either dynamically perturbed by the infall of the Magellanic system or was formerly a member of the Magellanic Clouds and was later stripped by the Milky Way. Our results showcase the potential of an all-sky search for low-metallicity targets with \textit{Gaia} XP and confirm that the methodology described here is an useful ``treasure map'' for finding additional UMP stars.

\end{abstract}


\section{Introduction} \label{sec:intro}

\setcounter{footnote}{10}
With the realization that ``weak-lined'' stars are characterized by high velocity dispersion \citep[][]{Roman1950}, many programs have searched in the Milky Way's halo for the lowest-metallicity stars \citep[][]{beers2005}. These metal-poor stars are surviving witnesses to the earliest episodes of chemical enrichment, 
which are inaccessible through other probes \citep[][]{frebel2015}. Specifically, the so-called ``ultra metal-poor'' (UMP) stars ($\feh < -4$\footnote{${\rm [X/H]} = \log_{10}\left( N_{\rm X}/N_{\rm H} \right)_\star - \log_{10}\left( N_{\rm X}/N_{\rm H} \right)_\odot$ for a given star $\star$ and element X.}) are presumed to be direct descendants of the first (Population III) stars, and their chemical abundances should preserve the yields of the first metal-free supernovae \citep[e.g.,][]{Ishigaki2018}. For comparison, even with very high-redshift data from \textit{JWST} (out to $z \sim 10$), measured metallicities in directly observed galaxies rarely reach $1/100$th the solar level \citep[e.g.,][]{Curti2024mzrHigh-z}, highlighting the key importance of local UMP stars in the Galactic halo. However, being able to study the high-$z$ Universe in detail with nearby UMP stars comes with the challenge that these are exquisitely rare, with less than 50 such objects known to this day 
\citep[e.g.,][]{bonifacio2025rev}.

The most prominent feature of Galactic metal-poor stars (here defined as $\feh \lesssim -2$) is that they are typically quite enhanced in carbon relative to iron \citep[${\rm [C/H] - [Fe/H] = [C/Fe] \gtrsim +0.7}$;][]{aoki2007}. \citet[][]{rossi1999} first recognized that the fraction of these carbon-enhanced metal-poor stars (CEMP) increases with decreasing metallicity. Indeed, nearly all stars with ${\rm [Fe/H]\leq-4.5}$ are of the carbon-enhanced kind \citep[][]{placco2014Carbon, Arentsen2022cemp}. For example, the most iron-deficient star known, \kellerstar, \citep[with $\feh < -7$;][]{keller2014} has an . carbon-to-iron ratio  ($\rm[C/Fe] > +4.5$). 

For quite some time, the only potential exception to this behavior has been the UMP star discovered by \citet[][]{caffau2011}, \caffaustar, with $\feh = -4.73$ and an upper limit of $\rm[C/Fe] < +0.91$ \citep[][]{caffau2024}. This carbon abundance limit is much lower than that of other stars at the same metallicity and it leaves room for \caffaustar to potentially not be a CEMP star. Consequently, and given reasonable assumptions about the nitrogen and oxygen abundances (see \citealt{caffau2011}), its total metal mass fraction is $\sim$10$\times$ lower than that of \kellerstar \citep[][]{keller2014}. Recently, additional non-CEMP UMP have been discovered both in the Milky Way \citep[][]{starkenburg2018ump, placco2021ump} and its satellites \citep[][]{Skuladottir2021UMPsculptor, chiti2024lmc}. These discoveries point to the chemical diversity of UMP stars and, hence, the heterogeneity in the properties of the first stars and supernovae themselves \citep[e.g.,][]{Heger2010}.

\renewcommand{\arraystretch}{1.0}
\setlength{\tabcolsep}{0.3em}
\begin{table*}[ht!]
\centering
\caption{Observational Data for \textit{Gaia} DR3  5262850721755411072
}
\label{tabelao}
\begin{tabular}{>{\normalsize}l >{\normalsize}l >{\normalsize}l >{\normalsize}l >{\normalsize}l >{\normalsize}l}
\hline
\hline
Quantity & Symbol & Value & Unit & Source & Reference\\ %
\hline
\hline
R.A. & $\alpha$ (J2000) & 108.91087 & degree & \textit{Gaia} DR3 & \citet[][]{GaiaDR3} \\ 
Decl. & $\delta$ (J2000) & $-$73.58142 & degree & \textit{Gaia} DR3 & \citet[][]{GaiaDR3}\\ 
Galactic longitude & $\ell$ & 284.97496 & degree & \textit{Gaia} DR3 & \citet[][]{GaiaDR3} \\ 
Galactic latitude & $b$ & $-$24.25953 & degree & \textit{Gaia} DR3 & \citet[][]{GaiaDR3}\\ 
Parallax & $\varpi$ & $-$0.0064$\pm$0.0172 & mas & \textit{Gaia} DR3 & \citet[][]{GaiaDR3}\\ 
Proper motion ($\alpha$) & $\mu_{\alpha,*}$\footnote{$\mu_{\alpha,*} = \mu_{\alpha} \cos \delta$} & 3.767$\pm$0.022 & mas yr$^{-1}$ & \textit{Gaia} DR3 & \citet[][]{GaiaDR3}\\ 
Proper motion ($\delta$) & $\mu_\delta$ & 2.627$\pm$0.022 & mas yr$^{-1}$ & \textit{Gaia} DR3 & \citet[][]{GaiaDR3}\\ 

\hline 
Mass & $M_\star$ & 0.78$\pm$0.10 & $M_\odot$ & Assumed & This work \\ 

\hline
$B$ magnitude & $B$ & 16.467 & mag & APASS\footnote{The AAVSO Photometric All-Sky Survey} DR9 & \citet[][]{apassdr9} \\
$V$ magnitude & $V$ & 15.362 & mag & APASS DR9 & \citet[][]{apassdr9} \\

$G$ magnitude & $G$ & 14.995 & mag & \textit{Gaia} DR3 & \citet[][]{GaiaDR3}\\ 
$BP$ magnitude & $BP$ & 15.648 & mag & \textit{Gaia} DR3 & \citet[][]{GaiaDR3}\\ 
$RP$ magnitude & $RP$ & 14.203 & mag & \textit{Gaia} DR3 & \citet[][]{GaiaDR3}\\ 

$J$ magnitude & $J$ & 13.161 & mag & 2MASS & \citet[][]{2MASS} 
\\
$H$ magnitude & $H$ & 12.579 & mag & 2MASS & \citet[][]{2MASS} \\
$K$ magnitude & $K$ & 12.466 & mag & 2MASS & \citet[][]{2MASS} \\

$W1$ magnitude & $W1$ & 12.311 & mag & AllWISE\footnote{Wide-field Infrared Survey Explorer} & \citet[][]{AllWISEcat} \\
$W2$ magnitude & $W2$ & 12.272 & mag & AllWISE & \citet[][]{AllWISEcat} \\

$g$ magnitude & $g$ & 15.682 & mag & SkyMapper DR4 & \citet[][]{SkyMapperDR4} \\
$r$ magnitude & $r$ & 15.053 & mag & SkyMapper DR4 & \citet[][]{SkyMapperDR4} \\
$i$ magnitude & $i$ & 14.550 & mag & SkyMapper DR4 & \citet[][]{SkyMapperDR4} \\
$z$ magnitude & $z$ & 14.304 & mag & SkyMapper DR4 & \citet[][]{SkyMapperDR4} \\

\hline 
Color excess & $E(B-V)$ & 0.1622 & mag & IRSA\footnote{Infrared Science Archive (\url{https://irsa.ipac.caltech.edu/})} & \citet[][]{Schlafly2011} \\ 
Extinction ($V$) & $A_V$ & 0.503 & mag & Equation \ref{extinct1} &  \citet[][]{Schlafly2011} \\ 
Extinction ($G$) & $A_G$ & 0.397$\pm$0.005 & mag & Equation \ref{extinct2} & \citet[][]{WangChen2019extinctions} \\ 
Bolometric correction & ${\rm BC}(G)$ & $-$0.568$\pm$0.022 & mag & \texttt{bcutil}\footnote{\url{https://github.com/casaluca/bolometric-corrections/tree/master}} & \citet[][]{CasagrandeVandenBerg2018gaia} \\

\hline 
Line-of-sight velocity & $v_{\rm los}$ & $+$428.7$\pm$2.0 & km\,s$^{-1}$ & MIKE/Magellan & This work \\

\hline 
Effective temperature & $\teff$ & 4596$\pm$65 & K & color--\teff--\feh\footnote{\citet[][]{mucciarelli2021calib}} & This work \\
&  & 4742 & K & Data driven & \citet[][]{andrae2023gaiaXP} \\
&  & 4803$^{+161}_{-51}$ & K & \texttt{StarHorse} & \citet[][]{Anders2022starhorseEDR3} \\

Log of surface gravity & $\logg$ & 0.88$\pm$0.15 & [cgs] & Isochrone\footnote{\citet[][]{YaleYonsei2001, YaleYonsei2004}} 
& This work \\
&  & 1.38 & [cgs] & Data driven & \citet[][]{andrae2023gaiaXP} \\
&  & 1.52$_{-0.25}^{+0.18}$ & [cgs] & \texttt{StarHorse} & \citet[][]{Anders2022starhorseEDR3} \\

Microturbulence velocity & \vturb & 2.27$\pm$0.10 & km\,s$^{-1}$ & \logg--\vturb\footnote{\citet[][quadratic fit by \citealt{AlexJi2023ret2}]{barklem2005}} & This work \\

Metallicity & \feh & $-$4.82$\pm$0.25 & dex & MIKE/Magellan & This work \\
& & $-$3.5 & dex & Data driven & This work \\
&  & $-$2.9 & dex & Data driven & \citet[][]{andrae2023gaiaXP} \\
&  & ${<}-$2.9 & dex & Model grid & M. K. Mardini et al., in prep. \\

\hline 
Distance modulus & $\mu$ & 16.9$\pm$0.4 & mag & Equation \ref{distlogg} & This work \\
Heliocentric distance & $d_{\rm helio}$ & 24.1$^{+4.9}_{-4.1}$ & kpc & $10^{[(\mu + 5)/5] - 3}$ & This work \\ 
& & 15.8$^{+3.4}_{-4.2}$ & kpc & \texttt{StarHorse} & \citet[][]{Anders2022starhorseEDR3} \\ 

\hline
\end{tabular}
\end{table*}

Clearly, identifying additional UMP stars is paramount to the quest to constrain the properties of the first supernovae. Current surveys are now optimized to discover the most metal-poor stars by using narrow-band photometry around metallicity-sensitive features of stellar spectra, notably the \ion{Ca}{2} K and H lines at $\lambda = 3933\,{\rm \AA}$ and $\lambda = 3969\,{\rm \AA}$, respectively. 
Then, high-resolution spectroscopy ($\mathcal{R}\gtrsim20k$) can be used to confirm the metallicities of the candidates and/or derive detailed elemental abundances. Some of these initiatives include the Pristine survey \citep[][and \citealt{aguado2019} for follow-up]{Starkenburg2017PRISTINE}, the Southern Photometric Local Universe Survey \citep[S-PLUS;][and see \citealt{Whitten2021splus}, \citealt{placco2022splus}, \citealt{splusSHORTSdr1}]{splus}, and the Javalambre Photometric Local Universe Survey \citep[J-PLUS;][\citealt{Whitten2019jplus}, \citealt{Galarza2022jplus}]{jplus}. The newest addition to this list is the Mapping the Ancient Galaxy in CaHK (MAGIC) survey (\citealt[][]{barbosa2025scl, placco2025magic}, A. Chiti et al., in prep.).

The success of the above-mentioned narrow-band photometric metallicity technique relies on imaging large areas of the sky to efficiently search for these rare low-metallicity stars. In this Letter, we showcase the power of an all-sky search based on the \textit{Gaia} mission's \citep[][]{GaiaMission} spectro-photometric data with the discovery of a UMP star with one of the lowest iron abundances ever measured ($\feh = -4.82 \pm 0.25$) and only an 
upper limit of carbon ($\cfe_{\rm cor} < +1.18$, see below). These properties make \textit{Gaia} DR3 5262850721755411072 \citep[][]{GaiaDR3}, hereafter ``\gdr'', one of the most metal-poor 
stars ever found.

\section{Data} \label{sec:data}

\begin{figure*}[pt!]
\centering
\includegraphics[width=2.0\columnwidth]{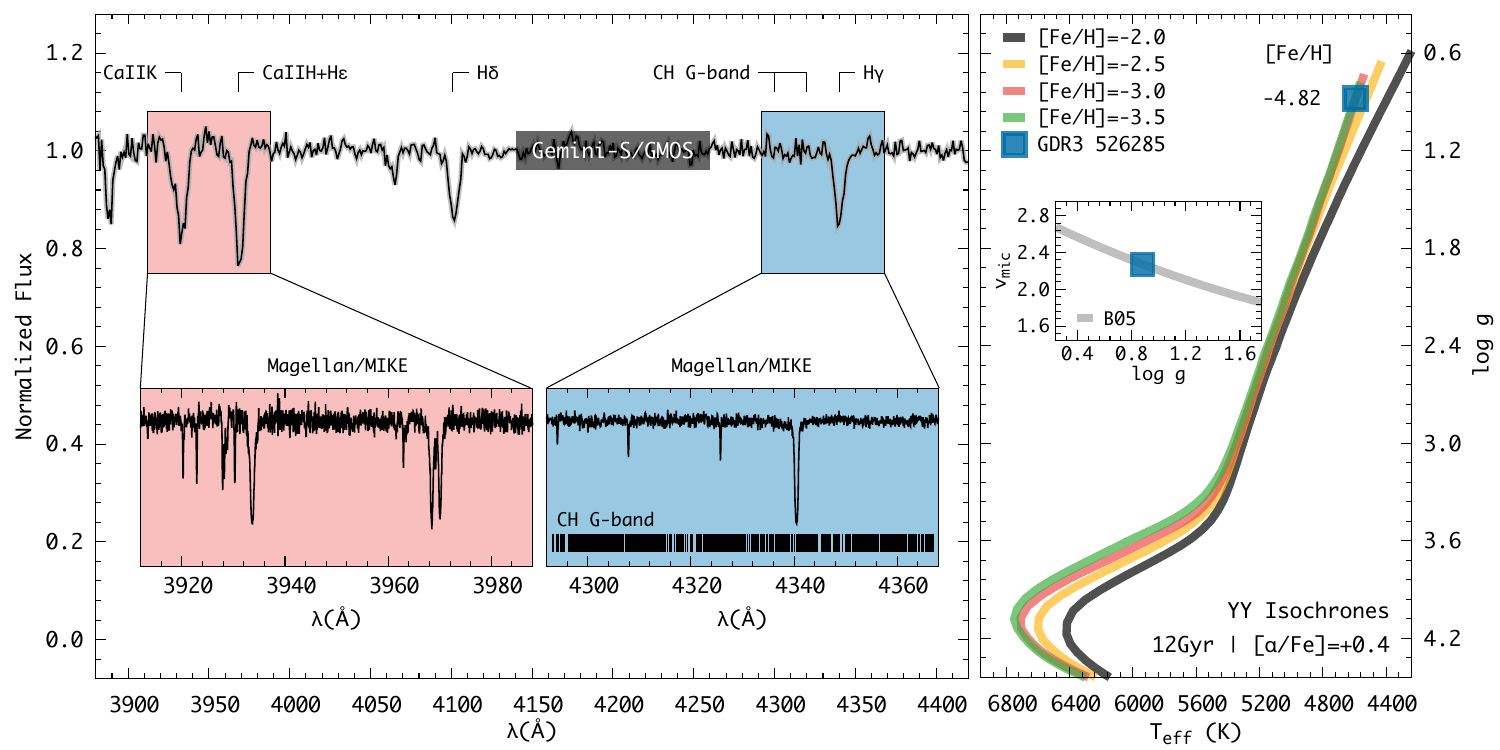}
\caption{Left: Normalized spectra of \gdr. The medium-resolution GMOS spectrum ($\mathcal{R} \sim 1500$; see text) is shown in the upper portion of this panel. Red and blue insets highlight spectral features of interest in our high-resolution MIKE spectrum ($\mathcal{R} \sim 35k$; Section \ref{obs}), namely \ion{Ca}{2} K and H ($3920 \lesssim \lambda/{\rm \AA} \lesssim 3980$) and CH G band ($\sim$4300\,\AA), respectively. Right: 12\,Gyr, $\alpha$-enhanced Y$^2$ isochrones \citep[][]{YaleYonsei2001, YaleYonsei2004} in \teff vs. \logg space at, from right to left, $\feh = -2$ (black), $-2.5$ (yellow), $-3$ (orange), and $-3.5$ (green). The inset panel shows the quadratic \logg--\vturb relation from \citet[][fit by \citealt{AlexJi2023ret2}, gray line]{barklem2005}. Star \gdr is plotted as the blue square.
\label{spec+isoc}}
\end{figure*}


\begin{figure*}[pt!]
\centering
\includegraphics[width=2.1\columnwidth]{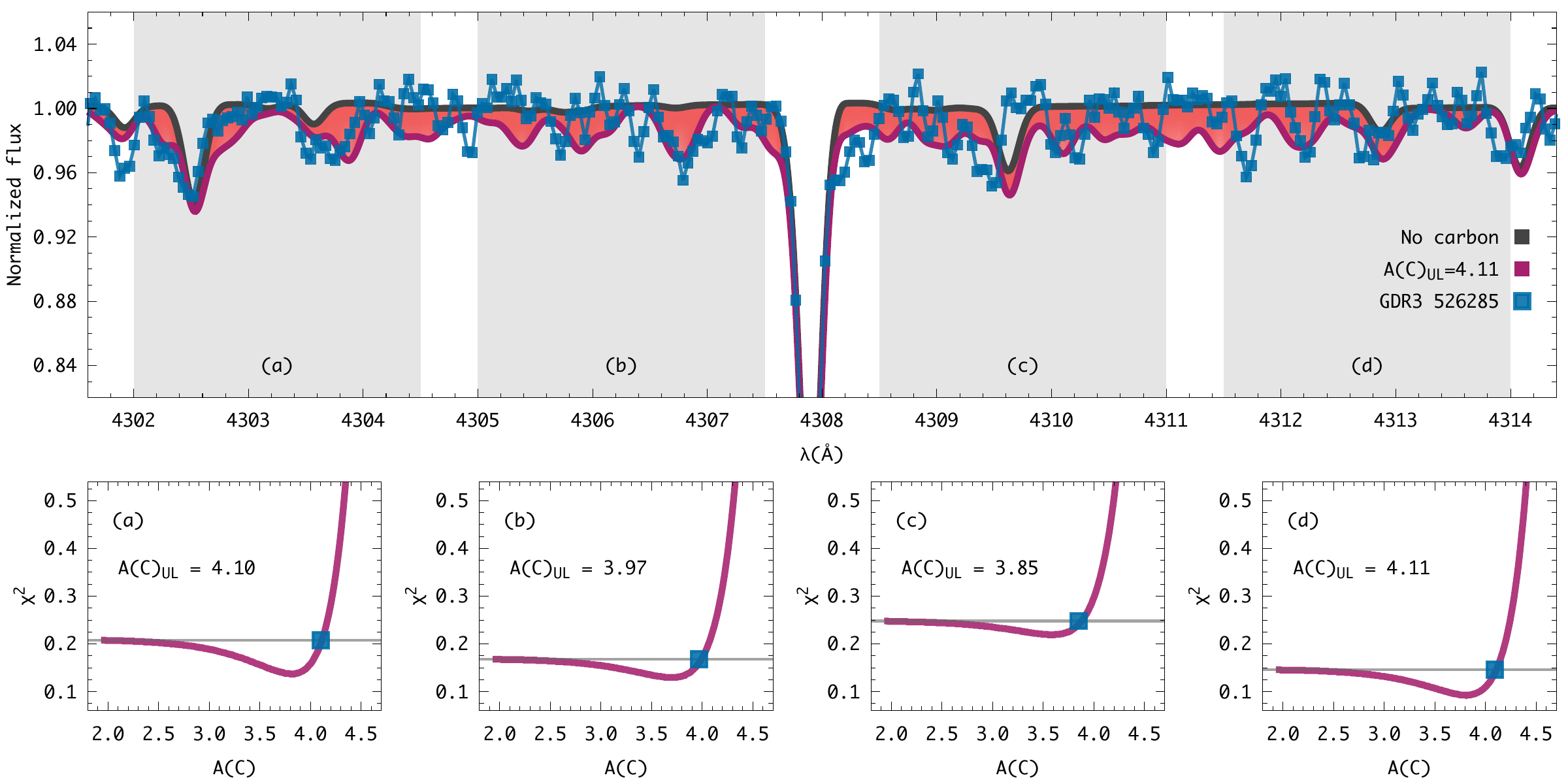}
\includegraphics[width=2.1\columnwidth]{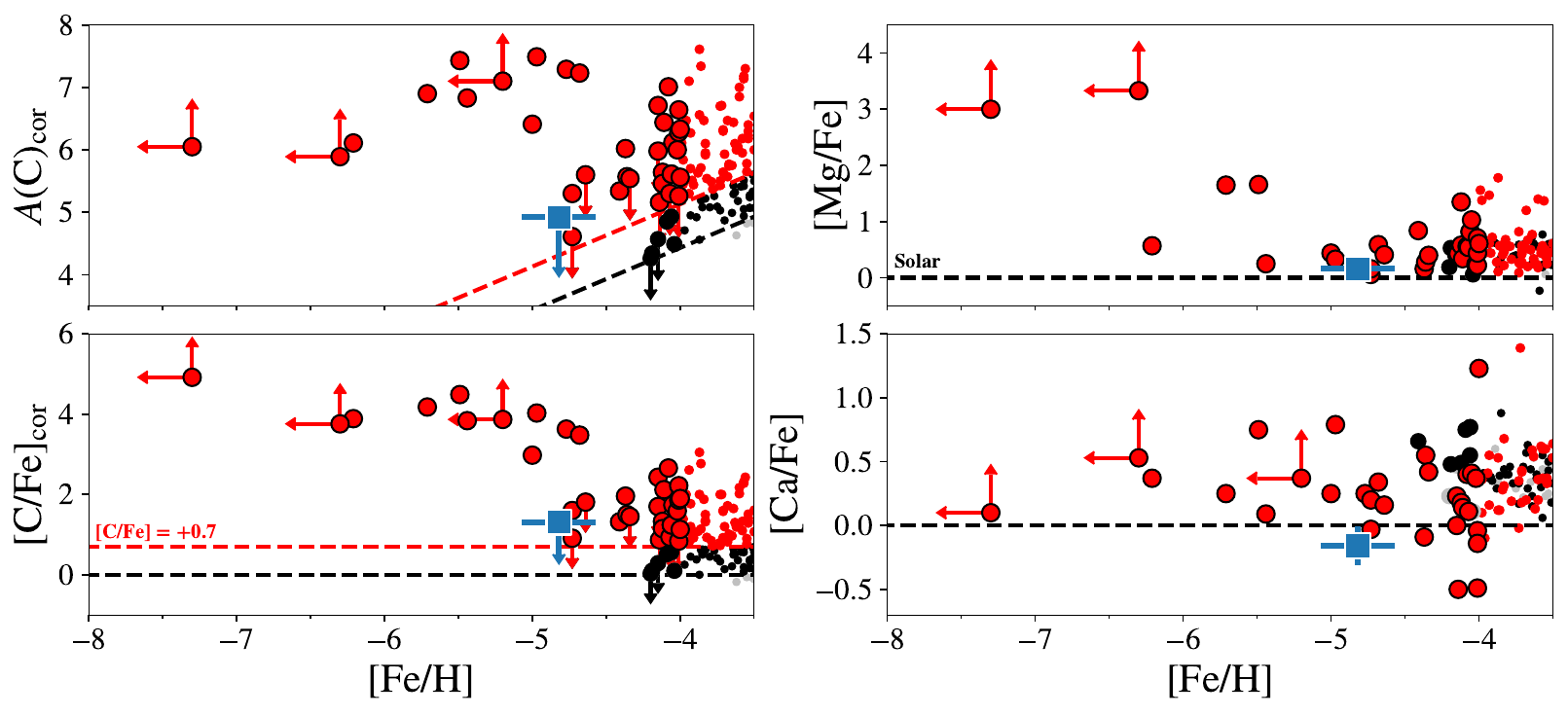}
\caption{Top panel: MIKE spectrum of \gdr around the CH G band (blue squares); $4300 \leq \lambda/{\rm \AA} \leq 4316$. 
Our final $A({\rm C}) = 4.11$ upper limit is represented by the solid magenta line. The difference between our $A({\rm C})$ upper limit and the no-carbon case is shown as the red region within magenta and black lines. Panels (a), (b), (c), and (d) illustrate different $\lambda$ windows used in our $\chi^2$ minimization method to derive the carbon-abundance upper limit (Section \ref{abunds}). Bottom panels: Key elemental abundance ratios that characterize UMP stars: $A({\rm C})_{\rm cor}$, [C/Fe]$_{\rm cor}$, [Mg/Fe], and [Ca/Fe] as a function of \feh, where the subscript ``cor'' indicates carbon abundances with applied evolutionary corrections \citep[][]{placco2014Carbon}. Small dots are non-UMP stars ($\feh > -4$) compiled from the SAGA database and JINAbase \citep[][]{SAGA_1, jina}, while the UMP stars are shown with larger circles. Symbols are colored according to stellar [C/Fe] ratios; subsolar in gray, $0 \leq {\rm [C/Fe]} < +0.7$ in black, and CEMP stars in red. Upper/lower limits are plotted with arrows. In the left-side carbon abundance panels, the red dashed lines mark this ${\rm [C/Fe]} = +0.7$ boundary for CEMP classification. Black dashed lines denote Solar levels in all panels. \gdr is the blue square (Table \ref{abund_table}).
\label{carbon+abunds}}
\end{figure*}

\subsection{Target selection}

We identified the UMP star \gdr in the \textit{Gaia} mission's third data release (DR3) of BP/RP (XP) low-resolution spectra \citep[$\mathcal{R} \approx 50$;][]{GaiaDR3, DeAngeli2022xp}. For the task, we utilize the \gxp-based catalog of candidate very metal-poor stars ($\feh < -2$) from \citet{Yao2023xp} as the starting point
. These authors constructed an \texttt{XGBoost} machine-learning \textit{classification} model (\citealt{XGBoost} for the algorithm) to flag likely low-metallicity stars in \gxp. We specifically explore their golden sample of $\approx$70,000 vetted red giant-branch (RGB) stars with effective temperatures (\teff) and surface gravity values (\logg) ideal for the search for stars with very weak absorption lines. To ensure that our UMP candidates were genuine RGB stars, we checked the stellar parameters from \citet[][also a data-driven method with \gxp]{andrae2023gaiaXP} before spectroscopic follow-up. 

To narrow our selection of the most promising UMP candidates, we trained an \texttt{XGBoost} \textit{regression} model to predict metallicities for the golden sample of low-metallicity RGB stars from \citet[][]{Yao2023xp} based on \gxp. For the training, we use the compilation of metallicities from high-resolution ($\mathcal{R} \geq 15k$) spectroscopic results available in either Stellar Abundances for Galactic Archaeology (SAGA) database or JINAbase \citep{suda2008, jina}\footnote{Exact training data for the \texttt{XGBoost} regression model: \citet{Cayrel2004}, \citet{Cohen2013}, \citet[][]{roederer2014}, \citet{Jacobson2015metalpoor}, \citet{holmbeck2020}, and \citet[][]{Li2022lamost}.} totaling $>$1200 stars. We combine our metallicity estimates with those from other methods in the literature, including the aforementioned \citet[][]{andrae2023gaiaXP} as well as an internal catalog of XP metallicities (M. K. Mardini et al., in prep.) that has previously been used to identify a UMP star in the Milky Way's disk \citep{mardini2024diskUMP} and another in the Large Magellanic Cloud \citep[LMC;][]{chiti2024lmc}. 

We selected \gdr for high-resolution spectroscopic follow-up due to its reddish color indicative of it being a cool star (\textit{Gaia} $BP - RP = 1.44$), with RGB-like stellar parameters in \citet[][$\teff = 4742\,{\rm K}$ and $\logg = 1.38$]{andrae2023gaiaXP}, and very low predicted metallicity from our regression model ($\feh_{\rm XP,Yao+23} = -3.5$). Our low metallicity value based on \gxp was corroborated by both the \citet[][$\feh_{\rm Andrae+23} = -2.9$]{andrae2023gaiaXP} and M. K. Mardini et al. (in prep., $\feh_{\rm CaHK} < -2.9$) catalogs. Lastly, \gdr is bright enough (\textit{Gaia} $BP = 15.6$) to be observed with high-resolution spectroscopy on a 6--8\,m-class telescope with reasonable exposure times 
that would allow us to detect absorption features of several different elements.

\subsection{Observations} \label{obs}

We obtained a high-resolution spectrum of \gdr on May 13, 2023 with the Magellan Inamori Kyocera Echelle \citep[MIKE;][]{Bernstein2003mike} instrument attached to the Magellan Clay 6.5\,m telescope located at Las Campanas Observatory, Chile. We gathered two exposures of 20\,min each using a 0\farcs7 slit and 2$\times$2 CCD binning, yielding a resolving power of $\mathcal{R} \sim 35k$ and $28k$ at the blue (wavelength $\lambda < 5000$\,\AA) and red ($\lambda > 5000$\,\AA) arms, respectively. The data were reduced with the standard MIKE-specific \texttt{CarPy} package\footnote{\url{https://code.obs.carnegiescience.edu/mike}} \citep[][]{carpy}. The final 1D spectrum reached $S/N \approx 20$ per pixel at $\lambda = 4000$\,\AA \ and $\approx$64 per pixel at $\lambda = 6500$\,\AA. Regions of interest in the MIKE spectrum are shown in the left panel of Figure \ref{spec+isoc}, namely \ion{Ca}{2} K and H at $3920 \lesssim \lambda/{\rm \AA} \lesssim 3980$ (red) and CH G band at $\sim$4300\,\AA \ (blue). Also shown is a medium-resolution ($\mathcal{R} \sim 1500$) spectrum of \gdr obtained with the Gemini (South) Multi-Object Spectrograph (GMOS) instrument \citep{GMOSPaper1, GMOSPaper2} at the Gemini South 8.1\,m telescope which shows a weak Ca\,K line and no visible CH absorption
\footnote{We recovered the GMOS spectrum from the Gemini Observatory Archive (\url{https://archive.gemini.edu/}); program ID GS-2023B-Q-107, PI G. Limberg. Two 750\,s exposures were obtained with the B600\,l\,mm$^{-1}$ (G5323) grating and a 1\farcs5 slit and 2$\times$2 spatial binning. We perform data reduction with the \texttt{DRAGONS} package \citep[][
]{DRAGONS2}.}.

With the reduced data at hand, we derive a line-of-sight velocity (\vlos) for \gdr by cross-correlating against a high-$S/N$ MIKE spectrum of metal-poor standard HD122563 \citep[$\feh = -2.75$;][]{Karovicova2020standards} using the Spectroscopy Made Harder \citep[\texttt{smhr};][]{smhr2025code} code\footnote{\url{https://github.com/andycasey/smhr}}. We find a heliocentric-corrected $\vlos = +428.7\kms$
. Due to slit centering and $\lambda$ calibration, MIKE spectra carry a $\sim$2\,$\kms$ systematic error floor at the resolution and $S/N$ of our data \citep[see][]{Ji2020MagLiteS, placco2025magic}.

\section{Methods} \label{sec:methods}

\subsection{Stellar parameters} \label{params}

We obtain stellar parameters for \gdr using a combination of multi-band photometric data and our MIKE spectrum. We apply color--\teff--\feh calibrations from \citet[][]{mucciarelli2021calib} assuming $\feh = -4$ and $G, BP, RP$ \textit{Gaia} DR3 photometry and $K$ magnitude from the Two Micron All Sky Survey \citep[2MASS;][]{2MASS}; Table \ref{tabelao} compiles relevant information. We compute 10$^4$ realizations of each of six possible color combinations accounting for the observed (Gaussian) uncertainties. Then, the final \teff value is taken as the weighted average between the medians and median absolute deviations of the resulting distributions. 

We point out that \gdr is located in a Galactic region of significant reddening at $E(B-V) = 0.1622$ \citep[][]{Schlafly2011}, which has a major impact on the inferred \teff. Hence, we consider the possibility of an overestimated reddening due to, e.g., the presence of the Magellanic gas stream \citep[][]{Nidever2010magstream} given the on-sky position of \gdr (see below). However, we notice that, even in the densest region of the Magellanic stream, within the bridge between the LMC and SMC\footnote{Small Magellanic Cloud.}, the \citet[][]{Schlafly2011} $E(B-V)$ is not as high. Therefore, we evaluate that the somewhat high reddening is very likely due to the location of \gdr near the Galactic plane (Galactic latitude $b = -24.2595$).
Still, we note that previous studies of high-reddening metal-poor stars \citep[e.g.,][]{Boniofacio2000metalpoor} performed their own corrections to the $E(B-V)$ values from, e.g., the \citet{Schlegel1998} dust map. Using equation 1 from \citet[][]{Boniofacio2000metalpoor} would here result in $E(B-V) = 0.158$, which is essentially the \citet[][]{Schlafly2011} value. Finally, considering a putative case of no extinction, the derived \teff with our method would then be $\approx$4320\,K. Such a cooler atmosphere would lead to an iron abundance of $\feh < -5$ for \gdr. As a result, we adopt the \citet{Schlafly2011} value which is, in fact, a conservative choice with regards to the low metallicity of our star.

We compare \gdr's \teff value with the most metal-poor models available in the Yonsei-Yale (Y$^2$) set of isochrones \citep[][]{YaleYonsei2001, YaleYonsei2004}. We particularly adopt a 12\,Gyr Y$^2$ track with $\feh = -3.5$ and $\alpha$ enhancement of $[\alpha/{\rm Fe}] = +0.4$. We interpolate this isochrone with \teff uncertainties to find \logg according to an RGB solution (right panel of Figure \ref{spec+isoc}). We note that \textit{Gaia} DR3 astrometric catalog has a negative parallax value for \gdr despite it being relatively bright at $G \approx 15$, reinforcing that this object is a distant giant rather than a nearby main-sequence star. 
We then calculate \gdr's microturbulence velocity ($\vturb$) from the empirical $\logg$--$\vturb$ quadratic relation from \citet[][]{AlexJi2023ret2} based on \citet[][]{barklem2005} data
. We underscore that a different $\logg$--$\vturb$ relation, e.g., with \citet[][]{roederer2014} parameters for RGB stars, yields \vturb values up to $\sim$0.5\,$\kms$ lower. Fixing all other stellar parameters, this would bring the \feh of \gdr to $-4.69$\,dex. Since this is within our formal 1$\sigma$ iron-abundance uncertainty (Table \ref{abund_table}), we consider our estimate of \vturb to be appropriate. 

With \teff, \logg, and \vturb fixed, we derive the iron abundance/metallicity for \gdr from the equivalent widths (Gaussian fits) of 30 \ion{Fe}{1} lines in the \texttt{smhr} environment. We use a \citet[][]{Castelli2003atmospheres} 1D plane-parallel model atmosphere with no overshooting and $\rm[M/H] = -4.0$. \texttt{smhr} runs the 2017 version of radiative transfer code \texttt{MOOG} \citep[][]{Sneden1973moog} with scattering \citep[][]{Sobeck2011}\footnote{\url{https://github.com/alexji/moog17scat}} under the assumption of local thermodynamic equilibrium (LTE). The final \feh, and other atmospheric parameters (\teff, \logg, \vturb), are shown in Table \ref{tabelao}.

For our kinematic analysis, we require an estimate of \gdr's heliocentric distance (\dhelio). We employ the fundamental relation
\begin{linenomath}
\begin{dmath}
\logg = 4.44 + \log M_\star + 4 \log \left( \dfrac{\teff}{5780\,{\rm K}} \right) + + 0.4(G_0 - \mu + {\rm BC}(G) - M_{\rm bol, \odot})
\label{distlogg}
\end{dmath}
\end{linenomath}
to derive a distance modulus ($\mu$), where $M_\star = 0.78\pm 0.1\, \msun$ is the assumed mass of \gdr \citep[following][]{barbosa2025scl}, $G_0$ is the extinction-corrected \textit{Gaia} $G$ magnitude, ${\rm BC}(G)$ is the corresponding bolometric correction \citep[][]{CasagrandeVandenBerg2018gaia}, and $M_{{\rm bol}, \odot} = 4.75$ is the bolometric magnitude of the Sun \citep[see][]{venn2017graces, roederer2018hd222925, Ji2020streams}. To obtain $G_0$, we utilize the extinction law from \citet[][]{WangChen2019extinctions} with $R_V = 3.1$:
\begin{equation}
A_V = R_V \cdot E(B-V),
\label{extinct1}    
\end{equation}
\begin{equation}
A_G = (0.789 \pm 0.005) \cdot A_V,
\label{extinct2}    
\end{equation}
\begin{equation}
G_0 = G - A_G,
\label{extinct3}    
\end{equation}
where $A_V$ and $A_G$ are extinction values in $V$ and $G$ bands, respectively, and $E(B-V)$ is the \citet[][]{Schlafly2011} color excess. We list the resulting $\mu$ and $d_{\rm helio}$, as well as other used quantities, in Table \ref{tabelao}. 

\subsection{Chemical abundances and upper limits} \label{abunds}

Abundances for chemical species beyond \ion{Fe}{1} are also determined with the \texttt{MOOG} 2017 version within \texttt{smhr}. The molecular and atomic line list is taken from the \texttt{linemake}\footnote{\url{https://github.com/vmplacco/linemake}} compilation \citep[][]{Placco2021linemake}. We default to equivalent widths (Gaussian fits) to measure abundances for \ion{Na}{1}, \ion{Mg}{1}, \ion{Ca}{1}, \ion{Sc}{2}, \ion{Ti}{2}, \ion{Cr}{1}, and \ion{Ni}{1}. For the CH band as well as for blended features, i.e., \ion{Al}{1} and \ion{Si}{1}, we use the spectral synthesis technique
. The number of lines detected for each element/ion ($N$) and final abundances and abundance ratios are in Table \ref{abund_table} assuming the \citet[][]{asplund2009} solar atmospheric composition. For species where $N =1$, we assume an uncertainty floor of 0.15\,dex.

Figure~\ref{carbon+abunds} shows the determination of the upper limit on carbon, using the method outlined in \citet{placco2024}. Briefly, we define the \ac upper limit as the abundance for which the $\chi^2$ matches the value for a spectrum without any carbon on the opposite side of the global minimum. The top panel shows the \gdr spectrum (blue squares and line, smoothed using a moving average of size 3), compared to a synthetic spectrum with no carbon (black line). The magenta line represents the carbon upper limit of $A({\rm C}) = 4.11$, $\rm[C/Fe]=+0.50$. The bottom panels show the $\chi^2$ minimization technique, applied to four 2.5\,\AA \ sections within the carbon G band. The adopted upper limit is the highest out of the four individual values.

\subsection{Kinematics} \label{kin}

We compute Cartesian Galactocentric positions $\mathbf{r} = (X,Y,Z)$ and velocities $\mathbf{v} = (V_X, V_Y, V_Z)$ for \gdr from phase-space information in Table \ref{tabelao} using \texttt{Astropy} tools \citep[][v5.3.4]{astropy, astropy2018}. We adopt the distance and velocity of the Sun with respect to the Galactic center to be $\sqrt{X^2_\odot + Y^2_\odot} = R_\odot = 8.122\,{\rm kpc}$, with no vertical displacement from the Milky Way's plane, and $(V_R, V_\phi, V_Z)_\odot = (-12.9,  245.6, 7.8)\kms$ in cylindrical frame\footnote{$V_R = (XV_X + YV_Y)/R$ and $V_\phi = (XV_Y - YV_X)/R$}, respectively \citep[][]{GRAVITY2018, Drimmel2018sunVel}. 
With $\mathbf{r}$ and $\mathbf{v}$, we calculate the total orbital energy $E_{\rm tot} = \frac{1}{2}(V_X^2 + V_Y^2 + V_Z^2) + \Phi (\mathbf{r})$ for \gdr in the axisymmetric Galactic model potential $\Phi$ of \citet[][]{mcmillan2017}. We also use the angular-momentum vector $\mathbf{L} = (L_X, L_Y, L_Z)$\footnote{$L_X = Y V_Z - Z V_Y$, $L_Y = Z V_X - X V_Z$, $L_Z = X V_Y - Y V_X$} for our interpretations. Note that $\etot < 0$ signifies that a given object is formally bound to the model potential $\Phi$ and $RV_\phi = L_Z < 0$ denotes prograde motion
.

\begin{deluxetable}{lrrrrrr}[!t]
\tablewidth{0pc}
\tablecaption{Abundances for Individual Species \label{abund_table}}
\tablehead{
\colhead{Species}                     & 
\colhead{$\log\epsilon_{\odot}$\,(X)} & 
\colhead{$\log\epsilon$\,(X)}         & 
\colhead{$\mbox{[X/H]}$}              & 
\colhead{$\mbox{[X/Fe]}$}             & 
\colhead{$\sigma$}                    & 
\colhead{$N$}                         }
\startdata
C           & 8.43 &  $<$4.11 & $<-$4.32  & $<+$0.50  & \nodata & \nodata \\
C$_{\rm cor}$\vv   & 8.43 &  $<$4.79 & $<-$3.64  & $<+$1.18 & \nodata & \nodata \\
\ion{Na}{1} & 6.24 &     1.77 &  $-$4.47  &  $+$0.35  &    0.10 &       2 \\
\ion{Mg}{1} & 7.60 &     2.94 &  $-$4.66  &  $+$0.16  &    0.10 &       5 \\
\ion{Al}{1} & 6.45 &     1.24 &  $-$5.21  &  $-$0.39  &    0.10 &       2 \\
\ion{Si}{1} & 7.51 &     2.85 &  $-$4.66  &  $+$0.16  &    0.15 &       1 \\
\ion{Ca}{1} & 6.34 &     1.36 &  $-$4.98  &  $-$0.16  &    0.15 &       1 \\
\ion{Sc}{2} & 3.15 &  $-$1.65 &  $-$4.79  &  $+$0.03  &    0.15 &       1 \\
\ion{Ti}{2} & 4.95 &     0.21 &  $-$4.74  &  $+$0.09  &    0.10 &       4 \\
\ion{Cr}{1} & 5.64 &     0.29 &  $-$5.35  &  $-$0.53  &    0.15 &       1 \\
\ion{Fe}{1} & 7.50 &     2.68 &  $-$4.82  &     0.00  &    0.25 &      30 \\
\ion{Ni}{1} & 6.22 &     1.62 &  $-$4.60  &  $+$0.22  &    0.15 &       1 \\
\enddata
\tablenotetext{a}{Applying the evolutionary carbon correction ($\Delta\mbox{[C/Fe]}=+0.68$) from \citet{placco2014Carbon}.}
\end{deluxetable}


\begin{figure*}[pt!]
\centering
\includegraphics[width=2.1\columnwidth]{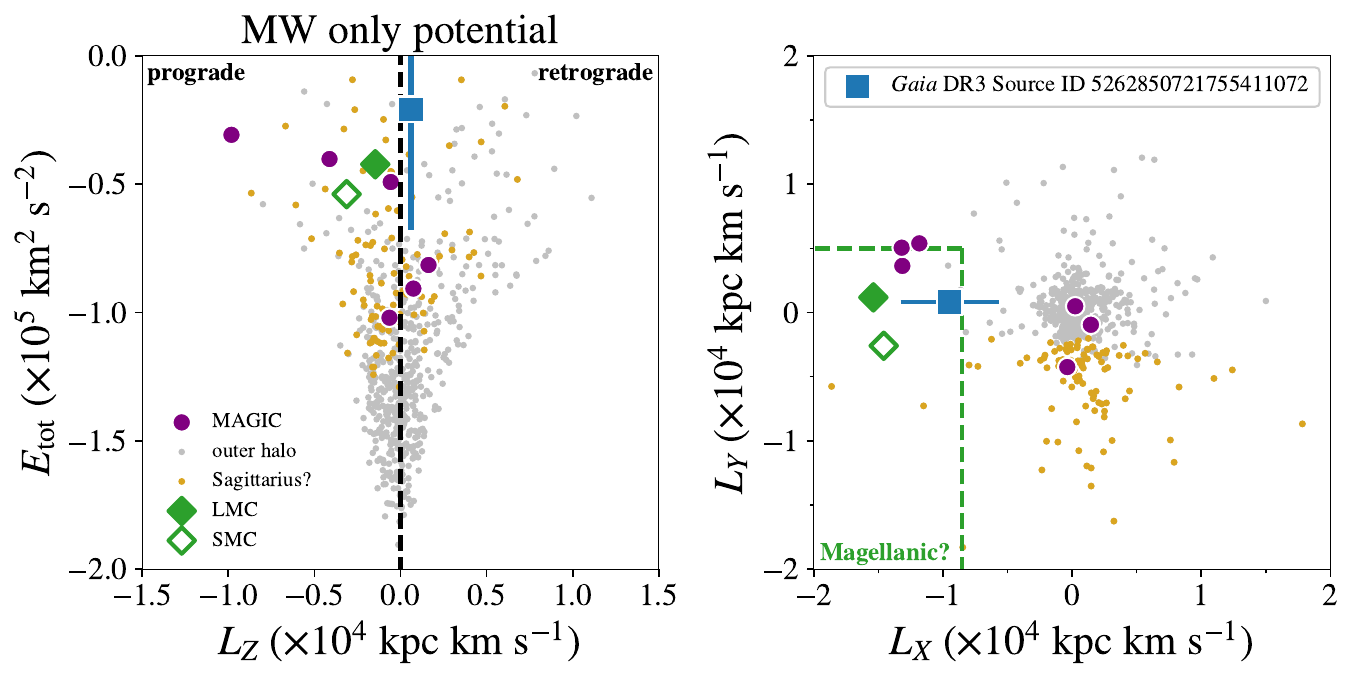}
\includegraphics[width=1.0\columnwidth]{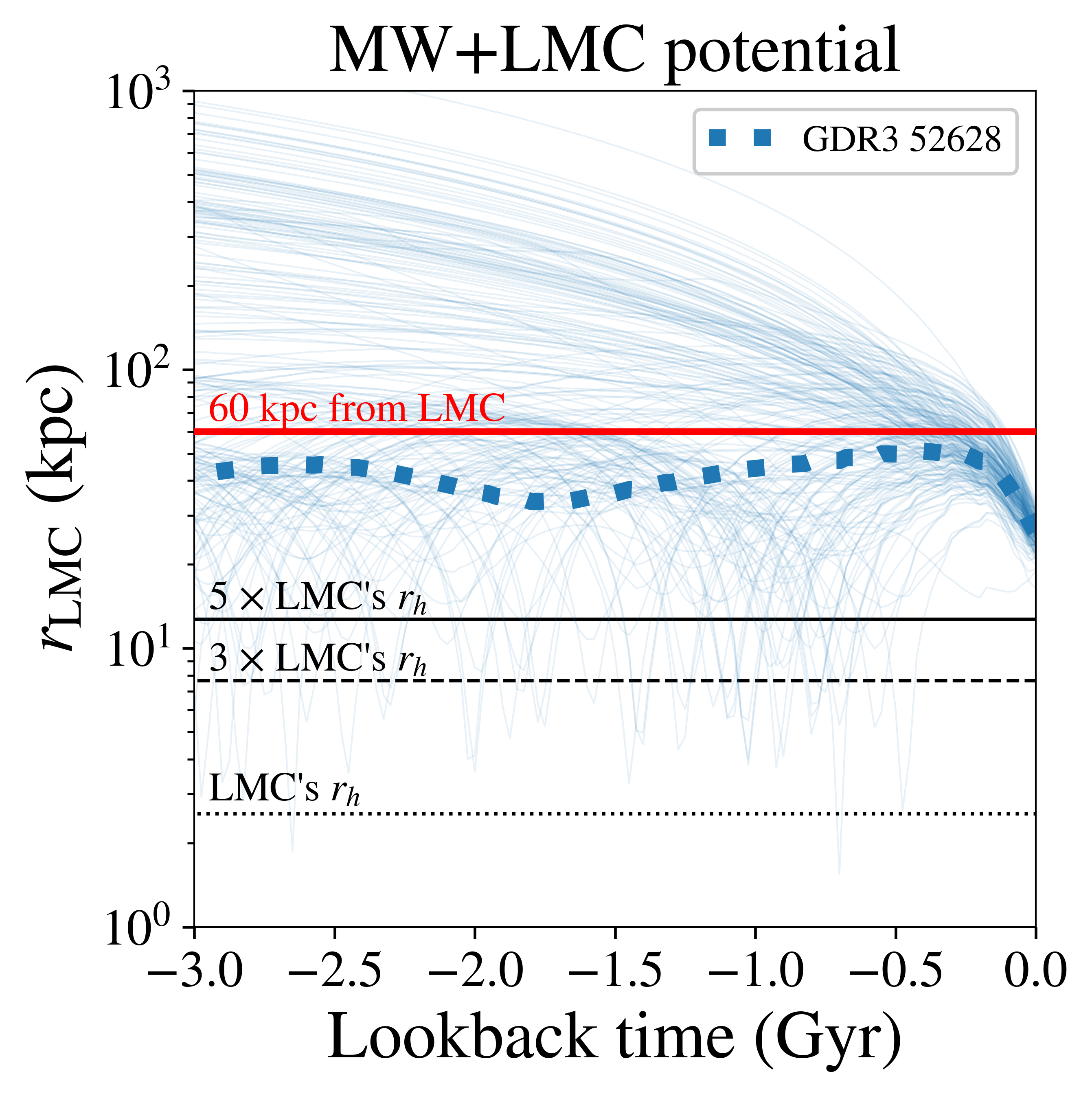}
\includegraphics[width=1.10\columnwidth]{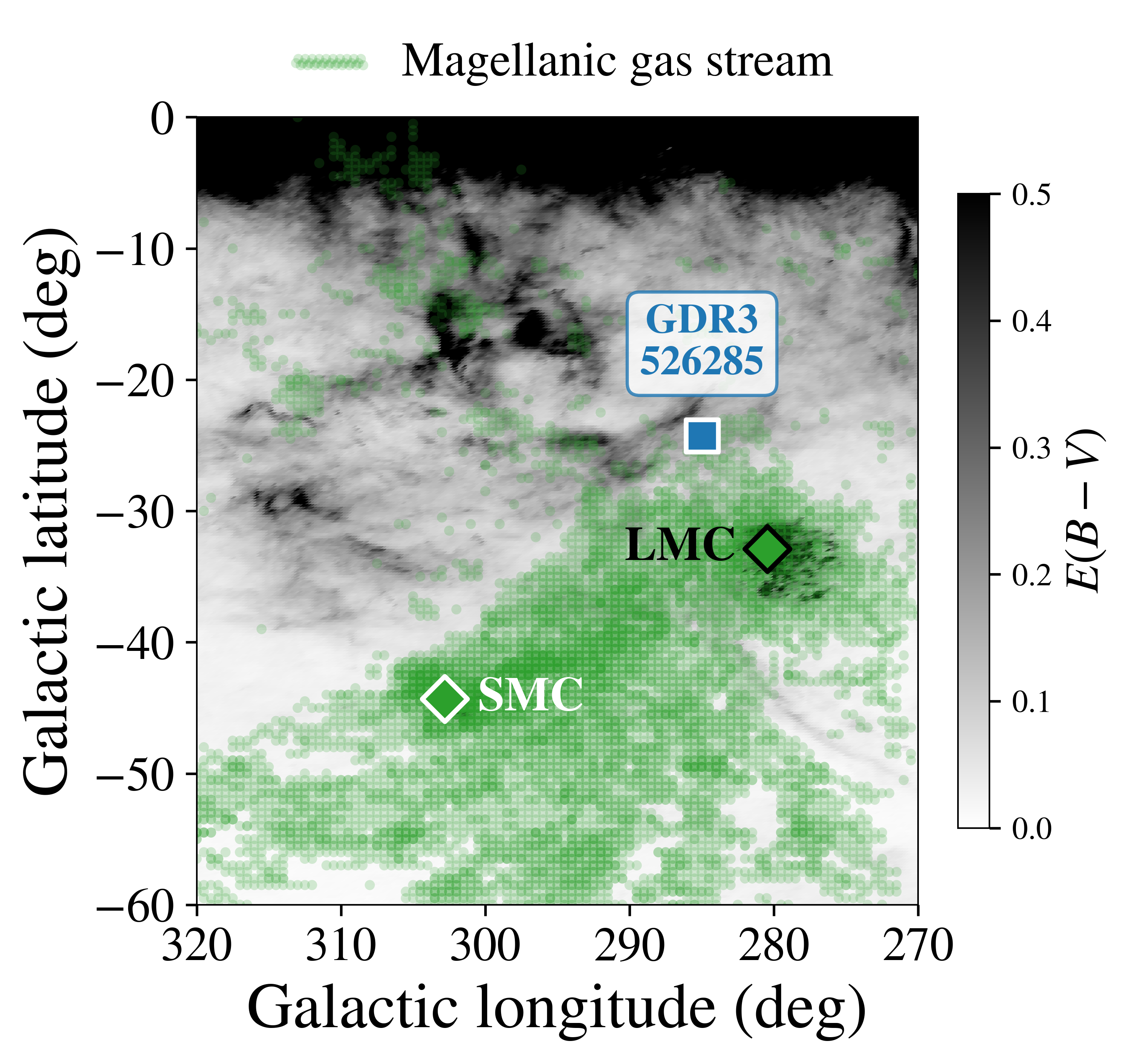}
\caption{Upper left: 
$(L_Z, E_{\rm tot})$, see text for discussion
. Upper right: $(L_X, L_Y)$. Gray and yellow dots are distant ($\geq$35\,kpc) low-metallicity ($\feh < -2.5$) halo stars \citep[][]{Limberg2023sgr}, where the latter are presumably associated with the Sagittarius stellar stream \citep[][]{Johnson2020sgr}. Purple circles are outer-halo extremely metal-poor stars ($\feh < -3$) from the MAGIC survey \citep{placco2025magic}. Green diamonds are the LMC (filled) and SMC (empty) based on phase-space information compiled by \citet[][\url{github.com/apace7/local_volume_database}]{pace2024lvdb}. Green dashed lines in $(L_X, L_Y)$ space delineate plausible association with the proposed Magellanic stellar stream \citep[][]{chandra2023stream}. In both upper panels, \gdr is represented by the blue square. 
Bottom left: 200 realizations of \gdr's orbit (thin blue lines, with mean trajectory as the dotted one) in a time-varying model potential that includes both the Milky Way and the infalling LMC \citep[][]{Vasiliev2021tango}. This figure depicts \gdr's distance to the LMC ($r_{\rm LMC}$) as a function of lookback time (backward orbit). Bottom right: On-sky location of \gdr, SMC, and LMC in Galactic coordinates. The distribution of $E(B-V)$ values from the \citet[][]{Schlegel1998} dust map is displayed as the grayscale background, with darker regions depicting higher reddening. The Magellanic gas stream \citep[][]{Nidever2010magstream} is overlaid in transparent green.
\label{starfit+kin}}
\end{figure*}

\section{Results} \label{results}

Our analysis based on multi-band photometry and high-resolution 
spectroscopy confirms that \gdr is a distant ($d_{\rm helio} = 24.1^{+4.9}_{-4.1}$) RGB star ($4596 \pm 65$\,K, $\logg = 0.88\pm 0.15$) in the Milky Way's halo. Comparing SkyMapper's DR4 \citep[][]{SkyMapperDR4} $g-r$ color (Table \ref{tabelao}) to a MESA Isochrones and Stellar Tracks \citep[MIST;][]{dotter2016, choi2016} 12.5\,Gyr, $\feh=-3$  solar-scaled isochrone, we find similar $\mu \approx 17.2$/$\dhelio \approx 27.5$\,kpc.

We note that the astro-photometric \dhelio value for \gdr derived by \citet[][$15.8^{+3.4}_{-4.2}$\,kpc]{Anders2022starhorseEDR3} with the \texttt{StarHorse} code \citep[][]{Santiago2016starhorse, Queiroz2018} is significantly lower than our own estimate. This \dhelio discrepancy might be due to a limitation of \texttt{StarHorse}'s adopted stellar model grid that only reaches $\rm[M/H] = -2.2$ at its metal-poor end; at a given \teff, lower-metallicity isochrones require \logg to be lower (see various isochrones in Figure \ref{spec+isoc}), which translates into a meaningful difference in the predicted $\mu$ (Equation \ref{distlogg}), hence \dhelio. Moreover, \textit{Gaia}'s negative parallax measurement is uninformative and \texttt{StarHorse}'s \dhelio estimate could be dominated by the assumed underlying Milky Way geometric prior, whereas, towards the on-sky location of \gdr, the outer Galactic halo is likely significantly perturbed by the presence of the LMC. To confirm this hypothesis, we calculate another \dhelio value with an exponentially decreasing space density prior following the method by \citet[][]{mardini2022atari}
. We find a \dhelio of $16.0\pm2.8$\,kpc, which is 1$\sigma$ compatible with the \texttt{StarHorse} value. We conclude that the \dhelio reported by \citet[][]{Anders2022starhorseEDR3} is biased low due to either one of these described effects or a combination thereof.

Beyond the extraordinarily low iron abundance of \gdr, perhaps the most remarkable feature of this UMP star is the relatively low upper limit on the [C/Fe] compared with other stars at a similar  [Fe/H]. 
There are two other known UMP stars at $\feh < -4.5$ with potentially similarly low carbon abundances, 
\caffaustar \citep{caffau2011} and Pristine\_221.8781$+$9.7844 \citep{starkenburg2018ump}. The latter has a relatively meaningless, high carbon abundance upper limit ($\rm[C/Fe] < +1.81$) compared to the former ($\rm[C/Fe] < +0.91$), when assuming the \citet[][]{asplund2009} solar chemical composition and LTE. 

The observed $\ac < 4.11$ upper limit we derive for \gdr is, in fact, nominally lower than the one for \caffaustar with $\ac < 4.61$ \citep[]{caffau2024}. However, \gdr is an evolved RGB star, and, if we compensate for the expected carbon depletion due to CN processing during stellar evolution \citep[$\Delta \cfe = +0.68$;][]{placco2014Carbon}, the final value we adopt is $\ac < 4.79$, leading to $\cfe < +1.18$ (bottom panels in Figure \ref{carbon+abunds}). Future spectroscopic observations with higher $S/N$ at $\sim$4300\,\AA \ will be needed to provide a more stringent constraint on the carbon abundance of \gdr. Moreover, departures from the 1D atmosphere and LTE assumptions affect cool metal-poor stars more profoundly, which would be the case for \gdr \citep[e.g.,][]{Asplund2005nlte}. Specifically for UMP ($\feh = -4$) RGB stars, \citet[][]{popa2023nlteCH} estimated corrections of about $+$0.2\,dex in [C/Fe] from the CH G band. Hence, future 3D non-LTE analysis of this star is also highly desirable.

The potential non-CEMP nature of \gdr along with its low iron abundance suggests that this stars has one of the lowest overall metal ($\text{atomic number} > 2$) mass fractions known, likely similar to that of \caffaustar. 
Consequently, \gdr could provide important constraints on the physics of early star formation in exceptionally metal-poor environments at redshift $z > 10$. For example, according to the critical threshold established by \citet[][]{frebel2007} for the cooling levels required for facilitating early low-mass star formation ($D_{\rm trans} \equiv \log_{10}{[10^{\rm [C/H]} + 0.3 \times 10^{\rm[O/H]}]}$), given our upper limit of $\rm[C/H]_{\rm cor} < -3.64$ and adopting oxygen enhancement of $\rm[O/Fe] = +0.6$, leads to $D_{\rm trans}<-3.6$. Even assuming $\rm[O/Fe] = +1.6$ only changes it to $D_{\rm trans}\lesssim-3.4$. These limits are close to the critical threshold of $D_{\rm trans,crit}\sim-3.5\pm0.2$, and future observations may well show this star to be below the critical value. As such, \gdr likely formed from dust-cooled gas \citep[][and references therein]{Chiaki2017} rather than fine-structure line cooling due to carbon and/or oxygen present in primordial gas. 

We estimate the overall metal mass fraction $Z$ for \gdr considering \textit{only} the elements/abundances from Table \ref{abund_table}. As in \citet[][]{caffau2011}, we assume both oxygen- and nitrogen-to-iron ratios to be $\rm[O \ or \ N/Fe] = +0.6$. For this calculation, we adopt the number ratio between helium and hydrogen atoms to be $1/12$. We find $Z_{\rm GDR3\_526585} \lesssim 1.0 \times 10^{-6}$ for \gdr. Using the same species and LTE values reported in \citet[][]{caffau2024} for \caffaustar leads to $Z_{\rm Caffau+24} \lesssim 9.2 \times 10^{-7}$. At least these two stars support a star formation mechanism in near-metal-free, dust-cooled environments.
Regarding the kinematics/dynamics, \gdr has a high orbital energy $E_{\rm tot} \gtrsim -0.5 \times 10^5\,{\rm km^2\,s^{-2}}$ (\citealt{mcmillan2017} potential) and an immense $X$-component of angular momentum $L_X = -9476^{+3458}_{-4140}\,{\rm kpc\,km\,s^{-1}}$. These are commensurate with its large \dhelio, proper motions, and \vlos (Table \ref{tabelao}). Similarly to high-\etot extremely metal-poor ($\feh < -3$) outer halo stars recently discovered in the MAGIC survey (\citealt{placco2025magic}; purple symbols in Figure \ref{starfit+kin}), it is plausible that the orbit of \gdr was perturbed by the infall of the LMC \citep[e.g.,][]{vasiliev2023lmc}. Another possibility is that \gdr was originally part of the Magellanic system, but has since been stripped by the Milky Way. If this latter hypothesis can be confirmed, \gdr might be linked to another UMP (non-CEMP) star in the LMC \citep[][]{chiti2024lmc}. The top right panel in Figure \ref{starfit+kin} shows, in $(L_X,L_Y)$ space, the cuts from \citet[][]{chandra2023stream} to isolate Magellanic stellar debris, and \gdr is located just within this region. The on-sky position of  \gdr in comparison to both SMC and LMC is presented in the bottom right panel of the same figure.

Given the hint from angular-momentum space that \gdr might have a Magellanic origin, we implement a Monte Carlo experiment similar to that of \citet[][]{chandra2023stream} to further evaluate this possibility. We integrate orbits for 3\,Gyr backward in a time-varying potential that includes both the Milky Way and the infalling LMC \citep[][]{Vasiliev2021tango}. In this model potential, the Milky Way stellar distribution is similar to that in \citet[][]{mcmillan2017}, but with a triaxial dark-matter halo
. The total assumed mass of the LMC is $M_{\rm LMC} = 1.5 \times 10^{11}\,\msun$. We sample 1,000 orbits for \gdr from the uncertainties in the observed phase-space quantities.
The results of this exercise are shown in the bottom left panel of Figure~\ref{starfit+kin}.
A realization is considered to have become bound to the LMC whenever it experiences a pericenter of $<$60\,kpc with respect to the LMC's center at $\geq$2\,Gyr ago (\citealt[][]{chandra2023stream} criterion; see the red solid line). The fraction of realizations where \gdr becomes bound to the LMC is 53.1\%, which reinforces a Magellanic association.

\section{Summary} \label{conclusions}

We have reported the discovery 
of a UMP star ($\feh = -4.82 \pm 0.25$) 
which has one of the lowest iron abundances ever measured. For reference, a search in both SAGA \citep[][]{suda2008} and the most up-to-date version of JINAbase\footnote{\url{https://github.com/Mohammad-Mardini/JINAbase-updated/}} \citep[][]{jina} reveals only $\sim$15 stars known at $\feh \lesssim -4.5$ \citep[e.g.,][]{Christlieb2004he0107-5240, frebel2005, Frebel2015ApJ, Frebel2019, hansen2014, Bonifacio2015, bonifacio2018topos, caffau2016, starkenburg2018ump, nordlander2019smss}, including the aforementioned \kellerstar \citep[][]{keller2014} and \caffaustar \citep[][]{caffau2011, caffau2024}. We note, however, that this exact number is susceptible to variations given uncertainties in \feh as well as 3D and non-LTE effects \citep[][]{NorrisYong2019nlte}.


We have identified this extraordinary star among the \textit{Gaia} DR3 low-resolution XP spectra, and combined multi-band photometry with high-resolution spectroscopy to derive stellar parameters and chemical abundances. Our analysis confirmed that \gdr is a distant ($\dhelio \approx 24\,{\rm kpc}$) RGB star ($\teff = 4596\,{\rm K}$, $\logg = 0.88$) in the Milky Way's halo. The star has an upper limit of $\cfe_{\rm cor} < +1.18$ (after applying the \citealt{placco2014Carbon} evolutionary correction) which suggests that it does not have a strong carbon overabundance as other stars with $\feh<-4.5$. This discovery highlights the potential of an all-sky search for low-metallicity targets with \textit{Gaia} XP and demonstrates that this strategy is a powerful ``treasure map'' for finding UMP stars.

Given the incredibly low abundances of \feh combined with no strong carbon enhancement, \gdr has an overall metal mass fraction of $Z_{\rm GDR3\_526585} \lesssim 1.0 \times 10^{-6}$ which is comparable to that of  \caffaustar \citep{caffau2011, caffau2024}. 
Future higher-\sn spectroscopic observations combined with 3D non-LTE modeling are of high interest to achieve a tighter upper limit on the carbon abundance, and perhaps a measurement of nitrogen, which has a major impact on the estimated overall metallicity. 
Finally, the kinematics of \gdr make it tentatively linked to the Magellanic system, either by being dynamically perturbed by its recent infall or as a former LMC star that has been tidally stripped by the Milky Way. We have performed a Monte Carlo experiment in a Milky Way$+$LMC potential and found that, for 53.1\% of the sampled orbits, \gdr indeed ended up bound to the LMC, pointing to a Magellanic origin.

%
%


\begin{acknowledgments}
We thank the referee for useful comments and suggestions that contributed to this manuscript. G.L. is indebted to all those involved with the multi-institutional \textit{Milky Way BR} Group for the weekly discussions. G.L. is particularly thankful to H\'elio Perottoni, who provided feedback on an early version of this paper, and Fred Anders, for assistance with \texttt{StarHorse} outputs and priors. 
G.L. acknowledges support from KICP/UChicago through a KICP Postdoctoral Fellowship. The work of V.M.P. is supported by NOIRLab, which is managed by the Association of Universities for Research in Astronomy (AURA) under a cooperative agreement with the U.S. National Science Foundation. A.C. is supported by a Brinson Prize Fellowship at KICP/UChicago. S.R. acknowledges partial financial support from FAPESP (proc. 2020/15245-2), CNPq (proc. 303816/2022-8), and CAPES. 
A.F. acknowledge support from NSF-AAG grant AST-2307436. 
This research is  partially based on observations obtained under program GS-2023B-Q-107 at the international Gemini Observatory.
  


This work has made use of data from the European Space Agency (ESA) mission {\it Gaia} (\url{https://www.cosmos.esa.int/gaia}), processed by the {\it Gaia} Data Processing and Analysis Consortium (DPAC, \url{https://www.cosmos.esa.int/web/gaia/dpac/consortium}). Funding for the DPAC has been provided by national institutions, in particular the institutions participating in the {\it Gaia} Multilateral Agreement.



\end{acknowledgments}

\newpage
\setlength{\bibsep}{0pt}
\bibliographystyle{aasjournal}
\footnotesize

\bibliography{bibliography.bib}{}

\begin{thebibliography}{}
\expandafter\ifx\csname natexlab\endcsname\relax\def\natexlab#1{#1}\fi
\providecommand{\url}[1]{\href{#1}{#1}}

\bibitem[{{Abohalima} \& {Frebel}(2018)}]{jina}
{Abohalima}, A., \& {Frebel}, A. 2018, \apjs, 238, 36

\bibitem[{{Aguado} {et~al.}(2019){Aguado}, {Youakim}, {Gonz{\'a}lez Hern{\'a}ndez}, {Allende Prieto}, {Starkenburg}, {Martin}, {Bonifacio}, {Arentsen}, {Caffau}, {Peralta de Arriba}, {Sestito}, {Garcia-Dias}, {Fantin}, {Hill}, {Jablonca}, {Jahandar}, {Kielty}, {Longeard}, {Lucchesi}, {S{\'a}nchez-Janssen}, {Osorio}, {Palicio}, {Tolstoy}, {Wilson}, {C{\^o}t{\'e}}, {Kordopatis}, {Lardo}, {Navarro}, {Thomas}, \& {Venn}}]{aguado2019}
{Aguado}, D.~S., {Youakim}, K., {Gonz{\'a}lez Hern{\'a}ndez}, J.~I., {et~al.} 2019, \mnras, 490, 2241

\bibitem[{{Anders} {et~al.}(2022){Anders}, {Khalatyan}, {Queiroz}, {Chiappini}, {Ard{\`e}vol}, {Casamiquela}, {Figueras}, {Jim{\'e}nez-Arranz}, {Jordi}, {Mongui{\'o}}, {Romero-G{\'o}mez}, {Altamirano}, {Antoja}, {Assaad}, {Cantat-Gaudin}, {Castro-Ginard}, {Enke}, {Girardi}, {Guiglion}, {Khan}, {Luri}, {Miglio}, {Minchev}, {Ramos}, {Santiago}, \& {Steinmetz}}]{Anders2022starhorseEDR3}
{Anders}, F., {Khalatyan}, A., {Queiroz}, A.~B.~A., {et~al.} 2022, \aap, 658, A91

\bibitem[{{Andrae} {et~al.}(2023){Andrae}, {Rix}, \& {Chandra}}]{andrae2023gaiaXP}
{Andrae}, R., {Rix}, H.-W., \& {Chandra}, V. 2023, \apjs, 267, 8

\bibitem[{{Aoki} {et~al.}(2007){Aoki}, {Beers}, {Christlieb}, {Norris}, {Ryan}, \& {Tsangarides}}]{aoki2007}
{Aoki}, W., {Beers}, T.~C., {Christlieb}, N., {et~al.} 2007, \apj, 655, 492

\bibitem[{{Arentsen} {et~al.}(2022){Arentsen}, {Placco}, {Lee}, {Aguado}, {Martin}, {Starkenburg}, \& {Yoon}}]{Arentsen2022cemp}
{Arentsen}, A., {Placco}, V.~M., {Lee}, Y.~S., {et~al.} 2022, arXiv e-prints, arXiv:2206.04081

\bibitem[{{Asplund}(2005)}]{Asplund2005nlte}
{Asplund}, M. 2005, \araa, 43, 481

\bibitem[{{Asplund} {et~al.}(2009){Asplund}, {Grevesse}, {Sauval}, \& {Scott}}]{asplund2009}
{Asplund}, M., {Grevesse}, N., {Sauval}, A.~J., \& {Scott}, P. 2009, \araa, 47, 481

\bibitem[{{Astropy Collaboration} {et~al.}(2013){Astropy Collaboration}, {Robitaille}, {Tollerud}, {Greenfield}, {Droettboom}, {Bray}, {Aldcroft}, {Davis}, {Ginsburg}, {Price-Whelan}, {Kerzendorf}, {Conley}, {Crighton}, {Barbary}, {Muna}, {Ferguson}, {Grollier}, {Parikh}, {Nair}, {Unther}, {Deil}, {Woillez}, {Conseil}, {Kramer}, {Turner}, {Singer}, {Fox}, {Weaver}, {Zabalza}, {Edwards}, {Azalee Bostroem}, {Burke}, {Casey}, {Crawford}, {Dencheva}, {Ely}, {Jenness}, {Labrie}, {Lim}, {Pierfederici}, {Pontzen}, {Ptak}, {Refsdal}, {Servillat}, \& {Streicher}}]{astropy}
{Astropy Collaboration}, {Robitaille}, T.~P., {Tollerud}, E.~J., {et~al.} 2013, \aap, 558, A33

\bibitem[{{Astropy Collaboration} {et~al.}(2018){Astropy Collaboration}, {Price-Whelan}, {Sip{\H{o}}cz}, {G{\"u}nther}, {Lim}, {Crawford}, {Conseil}, {Shupe}, {Craig}, {Dencheva}, {Ginsburg}, {Vand erPlas}, {Bradley}, {P{\'e}rez-Su{\'a}rez}, {de Val-Borro}, {Aldcroft}, {Cruz}, {Robitaille}, {Tollerud}, {Ardelean}, {Babej}, {Bach}, {Bachetti}, {Bakanov}, {Bamford}, {Barentsen}, {Barmby}, {Baumbach}, {Berry}, {Biscani}, {Boquien}, {Bostroem}, {Bouma}, {Brammer}, {Bray}, {Breytenbach}, {Buddelmeijer}, {Burke}, {Calderone}, {Cano Rodr{\'\i}guez}, {Cara}, {Cardoso}, {Cheedella}, {Copin}, {Corrales}, {Crichton}, {D'Avella}, {Deil}, {Depagne}, {Dietrich}, {Donath}, {Droettboom}, {Earl}, {Erben}, {Fabbro}, {Ferreira}, {Finethy}, {Fox}, {Garrison}, {Gibbons}, {Goldstein}, {Gommers}, {Greco}, {Greenfield}, {Groener}, {Grollier}, {Hagen}, {Hirst}, {Homeier}, {Horton}, {Hosseinzadeh}, {Hu}, {Hunkeler}, {Ivezi{\'c}}, {Jain}, {Jenness}, {Kanarek}, {Kendrew}, {Kern}, {Kerzendorf}, {Khvalko}, {King}, {Kirkby}, {Kulkarni},
  {Kumar}, {Lee}, {Lenz}, {Littlefair}, {Ma}, {Macleod}, {Mastropietro}, {McCully}, {Montagnac}, {Morris}, {Mueller}, {Mumford}, {Muna}, {Murphy}, {Nelson}, {Nguyen}, {Ninan}, {N{\"o}the}, {Ogaz}, {Oh}, {Parejko}, {Parley}, {Pascual}, {Patil}, {Patil}, {Plunkett}, {Prochaska}, {Rastogi}, {Reddy Janga}, {Sabater}, {Sakurikar}, {Seifert}, {Sherbert}, {Sherwood-Taylor}, {Shih}, {Sick}, {Silbiger}, {Singanamalla}, {Singer}, {Sladen}, {Sooley}, {Sornarajah}, {Streicher}, {Teuben}, {Thomas}, {Tremblay}, {Turner}, {Terr{\'o}n}, {van Kerkwijk}, {de la Vega}, {Watkins}, {Weaver}, {Whitmore}, {Woillez}, {Zabalza}, \& {Astropy Contributors}}]{astropy2018}
{Astropy Collaboration}, {Price-Whelan}, A.~M., {Sip{\H{o}}cz}, B.~M., {et~al.} 2018, \aj, 156, 123

\bibitem[{{Barbosa} {et~al.}(2025){Barbosa}, {Chiti}, {Limberg}, {Pace}, {Cerny}, {Rossi}, {Carlin}, {Stringfellow}, {Placco}, {Atzerberg}, {Carballo-Bello}, {Chaturvedi}, {Choi}, {Crnojevic}, {Drlica-Wagner}, {Ji}, {Kallivayalil}, {Mart{\'\i}nez-V{\'a}zquez}, {Medina}, {Noel}, {Sand}, {Vivas}, {Bom}, {Ferguson}, {Mutlu-Pakdil}, {Navabi}, {Riley}, {Sakowska}, \& {Zenteno}}]{barbosa2025scl}
{Barbosa}, F.~O., {Chiti}, A., {Limberg}, G., {et~al.} 2025, arXiv e-prints, arXiv:2504.03593

\bibitem[{{Barklem} {et~al.}(2005){Barklem}, {Christlieb}, {Beers}, {Hill}, {Bessell}, {Holmberg}, {Marsteller}, {Rossi}, {Zickgraf}, \& {Reimers}}]{barklem2005}
{Barklem}, P.~S., {Christlieb}, N., {Beers}, T.~C., {et~al.} 2005, \aap, 439, 129

\bibitem[{{Beers} \& {Christlieb}(2005)}]{beers2005}
{Beers}, T.~C., \& {Christlieb}, N. 2005, \araa, 43, 531

\bibitem[{{Bernstein} {et~al.}(2003){Bernstein}, {Shectman}, {Gunnels}, {Mochnacki}, \& {Athey}}]{Bernstein2003mike}
{Bernstein}, R., {Shectman}, S.~A., {Gunnels}, S.~M., {Mochnacki}, S., \& {Athey}, A.~E. 2003, in Society of Photo-Optical Instrumentation Engineers (SPIE) Conference Series, Vol. 4841, Instrument Design and Performance for Optical/Infrared Ground-based Telescopes, ed. M.~{Iye} \& A.~F.~M. {Moorwood}, 1694--1704

\bibitem[{{Bonifacio} {et~al.}(2025){Bonifacio}, {Caffau}, {Fran{\c{c}}ois}, \& {Spite}}]{bonifacio2025rev}
{Bonifacio}, P., {Caffau}, E., {Fran{\c{c}}ois}, P., \& {Spite}, M. 2025, arXiv e-prints, arXiv:2504.06335

\bibitem[{{Bonifacio} {et~al.}(2000){Bonifacio}, {Monai}, \& {Beers}}]{Boniofacio2000metalpoor}
{Bonifacio}, P., {Monai}, S., \& {Beers}, T.~C. 2000, \aj, 120, 2065

\bibitem[{{Bonifacio} {et~al.}(2015){Bonifacio}, {Caffau}, {Spite}, {Limongi}, {Chieffi}, {Klessen}, {Fran{\c{c}}ois}, {Molaro}, {Ludwig}, {Zaggia}, {Spite}, {Plez}, {Cayrel}, {Christlieb}, {Clark}, {Glover}, {Hammer}, {Koch}, {Monaco}, {Sbordone}, \& {Steffen}}]{Bonifacio2015}
{Bonifacio}, P., {Caffau}, E., {Spite}, M., {et~al.} 2015, \aap, 579, A28

\bibitem[{{Bonifacio} {et~al.}(2018){Bonifacio}, {Caffau}, {Spite}, {Spite}, {Sbordone}, {Monaco}, {Fran{\c{c}}ois}, {Plez}, {Molaro}, {Gallagher}, {Cayrel}, {Christlieb}, {Klessen}, {Koch}, {Ludwig}, {Steffen}, {Zaggia}, \& {Abate}}]{bonifacio2018topos}
---. 2018, \aap, 612, A65

\bibitem[{{Caffau} {et~al.}(2011){Caffau}, {Bonifacio}, {Fran{\c{c}}ois}, {Sbordone}, {Monaco}, {Spite}, {Spite}, {Ludwig}, {Cayrel}, {Zaggia}, {Hammer}, {Randich}, {Molaro}, \& {Hill}}]{caffau2011}
{Caffau}, E., {Bonifacio}, P., {Fran{\c{c}}ois}, P., {et~al.} 2011, \nat, 477, 67

\bibitem[{{Caffau} {et~al.}(2016){Caffau}, {Bonifacio}, {Spite}, {Spite}, {Monaco}, {Sbordone}, {Fran{\c{c}}ois}, {Gallagher}, {Plez}, {Zaggia}, {Ludwig}, {Cayrel}, {Koch}, {Steffen}, {Salvadori}, {Klessen}, {Glover}, \& {Christlieb}}]{caffau2016}
{Caffau}, E., {Bonifacio}, P., {Spite}, M., {et~al.} 2016, \aap, 595, L6

\bibitem[{{Caffau} {et~al.}(2024){Caffau}, {Bonifacio}, {Monaco}, {Steffen}, {Sbordone}, {Spite}, {Fran{\c{c}}ois}, {Gallagher}, {Ludwig}, \& {Molaro}}]{caffau2024}
{Caffau}, E., {Bonifacio}, P., {Monaco}, L., {et~al.} 2024, \aap, 691, A245

\bibitem[{{Casagrande} \& {VandenBerg}(2018)}]{CasagrandeVandenBerg2018gaia}
{Casagrande}, L., \& {VandenBerg}, D.~A. 2018, \mnras, 479, L102

\bibitem[{{Casey} {et~al.}(2025){Casey}, {Ji}, \& {Holmbeck}}]{smhr2025code}
{Casey}, A.~R., {Ji}, A., \& {Holmbeck}, E. 2025, {smhr: Automatic curve-of-growth analyses of high-resolution stellar spectra}, Astrophysics Source Code Library, record ascl:2502.025, ,

\bibitem[{{Castelli} \& {Kurucz}(2003)}]{Castelli2003atmospheres}
{Castelli}, F., \& {Kurucz}, R.~L. 2003, in Modelling of Stellar Atmospheres, ed. N.~{Piskunov}, W.~W. {Weiss}, \& D.~F. {Gray}, Vol. 210, A20

\bibitem[{{Cayrel} {et~al.}(2004){Cayrel}, {Depagne}, {Spite}, {Hill}, {Spite}, {Fran{\c{c}}ois}, {Plez}, {Beers}, {Primas}, {Andersen}, {Barbuy}, {Bonifacio}, {Molaro}, \& {Nordstr{\"o}m}}]{Cayrel2004}
{Cayrel}, R., {Depagne}, E., {Spite}, M., {et~al.} 2004, \aap, 416, 1117

\bibitem[{{Cenarro} {et~al.}(2019){Cenarro}, {Moles}, {Crist{\'o}bal-Hornillos}, {Mar{\'\i}n-Franch}, {Ederoclite}, {Varela}, {L{\'o}pez-Sanjuan}, {Hern{\'a}ndez-Monteagudo}, {Angulo}, {V{\'a}zquez Rami{\'o}}, {Viironen}, {Bonoli}, {Orsi}, {Hurier}, {San Roman}, {Greisel}, {Vilella-Rojo}, {D{\'\i}az-Garc{\'\i}a}, {Logro{\~n}o-Garc{\'\i}a}, {Gurung-L{\'o}pez}, {Spinoso}, {Izquierdo-Villalba}, {Aguerri}, {Allende Prieto}, {Bonatto}, {Carvano}, {Chies-Santos}, {Daflon}, {Dupke}, {Falc{\'o}n-Barroso}, {Gon{\c{c}}alves}, {Jim{\'e}nez-Teja}, {Molino}, {Placco}, {Solano}, {Whitten}, {Abril}, {Ant{\'o}n}, {Bello}, {Bielsa de Toledo}, {Castillo-Ram{\'\i}rez}, {Chueca}, {Civera}, {D{\'\i}az-Mart{\'\i}n}, {Dom{\'\i}nguez-Mart{\'\i}nez}, {Garzar{\'a}n-Calderaro}, {Hern{\'a}ndez-Fuertes}, {Iglesias-Marzoa}, {I{\~n}iguez}, {Jim{\'e}nez Ruiz}, {Kruuse}, {Lamadrid}, {Lasso-Cabrera}, {L{\'o}pez-Alegre}, {L{\'o}pez-Sainz}, {Ma{\'\i}cas}, {Moreno-Signes}, {Muniesa}, {Rodr{\'\i}guez-Llano}, {Rueda-Teruel}, {Rueda-Teruel},
  {Soriano-Lagu{\'\i}a}, {Tilve}, {Valdivielso}, {Yanes-D{\'\i}az}, {Alcaniz}, {Mendes de Oliveira}, {Sodr{\'e}}, {Coelho}, {Lopes de Oliveira}, {Tamm}, {Xavier}, {Abramo}, {Akras}, {Alfaro}, {Alvarez-Cand al}, {Ascaso}, {Beasley}, {Beers}, {Borges Fernandes}, {Bruzual}, {Buzzo}, {Carrasco}, {Cepa}, {Cortesi}, {Costa-Duarte}, {De Pr{\'a}}, {Favole}, {Galarza}, {Galbany}, {Garcia}, {Gonz{\'a}lez Delgado}, {Gonz{\'a}lez-Serrano}, {Guti{\'e}rrez-Soto}, {Hernandez-Jimenez}, {Kanaan}, {Kuncarayakti}, {Landim}, {Laur}, {Licandro}, {Lima Neto}, {Lyman}, {Ma{\'\i}z Apell{\'a}niz}, {Miralda-Escud{\'e}}, {Morate}, {Nogueira-Cavalcante}, {Novais}, {Oncins}, {Oteo}, {Overzier}, {Pereira}, {Rebassa-Mansergas}, {Reis}, {Roig}, {Sako}, {Salvador-Rusi{\~n}ol}, {Sampedro}, {S{\'a}nchez-Bl{\'a}zquez}, {Santos}, {Schmidtobreick}, {Siffert}, {Telles}, \& {Vilchez}}]{jplus}
{Cenarro}, A.~J., {Moles}, M., {Crist{\'o}bal-Hornillos}, D., {et~al.} 2019, A\&A, 622, A176

\bibitem[{{Chandra} {et~al.}(2023){Chandra}, {Naidu}, {Conroy}, {Bonaca}, {Zaritsky}, {Cargile}, {Caldwell}, {Johnson}, {Han}, \& {Ting}}]{chandra2023stream}
{Chandra}, V., {Naidu}, R.~P., {Conroy}, C., {et~al.} 2023, \apj, 956, 110

\bibitem[{{Chen} \& {Guestrin}(2016)}]{XGBoost}
{Chen}, T., \& {Guestrin}, C. 2016, arXiv e-prints, arXiv:1603.02754

\bibitem[{{Chiaki} {et~al.}(2017){Chiaki}, {Tominaga}, \& {Nozawa}}]{Chiaki2017}
{Chiaki}, G., {Tominaga}, N., \& {Nozawa}, T. 2017, \mnras, 472, L115

\bibitem[{{Chiti} {et~al.}(2024){Chiti}, {Mardini}, {Limberg}, {Frebel}, {Ji}, {Reggiani}, {Ferguson}, {Andales}, {Brauer}, {Li}, \& {Simon}}]{chiti2024lmc}
{Chiti}, A., {Mardini}, M., {Limberg}, G., {et~al.} 2024, Nature Astronomy, 8, 637

\bibitem[{{Choi} {et~al.}(2016){Choi}, {Dotter}, {Conroy}, {Cantiello}, {Paxton}, \& {Johnson}}]{choi2016}
{Choi}, J., {Dotter}, A., {Conroy}, C., {et~al.} 2016, \apj, 823, 102

\bibitem[{{Christlieb} {et~al.}(2004){Christlieb}, {Gustafsson}, {Korn}, {Barklem}, {Beers}, {Bessell}, {Karlsson}, \& {Mizuno-Wiedner}}]{Christlieb2004he0107-5240}
{Christlieb}, N., {Gustafsson}, B., {Korn}, A.~J., {et~al.} 2004, \apj, 603, 708

\bibitem[{{Cohen} {et~al.}(2013){Cohen}, {Christlieb}, {Thompson}, {McWilliam}, {Shectman}, {Reimers}, {Wisotzki}, \& {Kirby}}]{Cohen2013}
{Cohen}, J.~G., {Christlieb}, N., {Thompson}, I., {et~al.} 2013, \apj, 778, 56

\bibitem[{{Curti} {et~al.}(2024){Curti}, {Maiolino}, {Curtis-Lake}, {Chevallard}, {Carniani}, {D'Eugenio}, {Looser}, {Scholtz}, {Charlot}, {Cameron}, {{\"U}bler}, {Witstok}, {Boyett}, {Laseter}, {Sandles}, {Arribas}, {Bunker}, {Giardino}, {Maseda}, {Rawle}, {Rodr{\'\i}guez Del Pino}, {Smit}, {Willott}, {Eisenstein}, {Hausen}, {Johnson}, {Rieke}, {Robertson}, {Tacchella}, {Williams}, {Willmer}, {Baker}, {Bhatawdekar}, {Egami}, {Helton}, {Ji}, {Kumari}, {Perna}, {Shivaei}, \& {Sun}}]{Curti2024mzrHigh-z}
{Curti}, M., {Maiolino}, R., {Curtis-Lake}, E., {et~al.} 2024, \aap, 684, A75

\bibitem[{{Cutri} {et~al.}(2021){Cutri}, {Wright}, {Conrow}, {Fowler}, {Eisenhardt}, {Grillmair}, {Kirkpatrick}, {Masci}, {McCallon}, {Wheelock}, {Fajardo-Acosta}, {Yan}, {Benford}, {Harbut}, {Jarrett}, {Lake}, {Leisawitz}, {Ressler}, {Stanford}, {Tsai}, {Liu}, {Helou}, {Mainzer}, {Gettngs}, {Gonzalez}, {Hoffman}, {Marsh}, {Padgett}, {Skrutskie}, {Beck}, {Papin}, \& {Wittman}}]{AllWISEcat}
{Cutri}, R.~M., {Wright}, E.~L., {Conrow}, T., {et~al.} 2021, {VizieR Online Data Catalog: AllWISE Data Release (Cutri+ 2013)}, VizieR On-line Data Catalog: II/328. Originally published in: IPAC/Caltech (2013), ,

\bibitem[{{Davies} {et~al.}(1997){Davies}, {Allington-Smith}, {Bettess}, {Chadwick}, {Content}, {Dodsworth}, {Haynes}, {Lee}, {Lewis}, {Webster}, {Atad}, {Beard}, {Ellis}, {Hastings}, {Williams}, {Bond}, {Crampton}, {Davidge}, {Fletcher}, {Leckie}, {Morbey}, {Murowinski}, {Roberts}, {Saddlemyer}, {Sebesta}, {Stilburn}, \& {Szeto}}]{GMOSPaper1}
{Davies}, R.~L., {Allington-Smith}, J.~R., {Bettess}, P., {et~al.} 1997, in Society of Photo-Optical Instrumentation Engineers (SPIE) Conference Series, Vol. 2871, Optical Telescopes of Today and Tomorrow, ed. A.~L. {Ardeberg}, 1099--1106

\bibitem[{{De Angeli} {et~al.}(2022){De Angeli}, {Weiler}, {Montegriffo}, {Evans}, {Riello}, {Andrae}, {Carrasco}, {Busso}, {Burgess}, {Cacciari}, {Davidson}, {Harrison}, {Hodgkin}, {Jordi}, {Osborne}, {Pancino}, {Altavilla}, {Barstow}, {Bailer-Jones}, {Bellazzini}, {Brown}, {Castellani}, {Cowell}, {Delchambre}, {De Luise}, {Diener}, {Fabricius}, {Fouesneau}, {Fremat}, {Gilmore}, {Giuffrida}, {Hambly}, {Hidalgo}, {Holland}, {Kostrzewa-Rutkowska}, {van Leeuwen}, {Lobel}, {Marinoni}, {Miller}, {Pagani}, {Palaversa}, {Piersimoni}, {Pulone}, {Ragaini}, {Rainer}, {Richards}, {Rixon}, {Ruz-Mieres}, {Sanna}, {Sarro}, {Rowell}, {Sordo}, {Walton}, \& {Yoldas}}]{DeAngeli2022xp}
{De Angeli}, F., {Weiler}, M., {Montegriffo}, P., {et~al.} 2022, arXiv e-prints, arXiv:2206.06143

\bibitem[{{Demarque} {et~al.}(2004){Demarque}, {Woo}, {Kim}, \& {Yi}}]{YaleYonsei2004}
{Demarque}, P., {Woo}, J.-H., {Kim}, Y.-C., \& {Yi}, S.~K. 2004, \apjs, 155, 667

\bibitem[{{Dotter}(2016)}]{dotter2016}
{Dotter}, A. 2016, \apjs, 222, 8

\bibitem[{{Drimmel} \& {Poggio}(2018)}]{Drimmel2018sunVel}
{Drimmel}, R., \& {Poggio}, E. 2018, Research Notes of the American Astronomical Society, 2, 210

\bibitem[{{Frebel} {et~al.}(2015){Frebel}, {Chiti}, {Ji}, {Jacobson}, \& {Placco}}]{Frebel2015ApJ}
{Frebel}, A., {Chiti}, A., {Ji}, A.~P., {Jacobson}, H.~R., \& {Placco}, V.~M. 2015, \apjl, 810, L27

\bibitem[{{Frebel} {et~al.}(2019){Frebel}, {Ji}, {Ezzeddine}, {Hansen}, {Chiti}, {Thompson}, \& {Merle}}]{Frebel2019}
{Frebel}, A., {Ji}, A.~P., {Ezzeddine}, R., {et~al.} 2019, \apj, 871, 146

\bibitem[{{Frebel} {et~al.}(2007){Frebel}, {Johnson}, \& {Bromm}}]{frebel2007}
{Frebel}, A., {Johnson}, J.~L., \& {Bromm}, V. 2007, \mnras, 380, L40

\bibitem[{{Frebel} \& {Norris}(2015)}]{frebel2015}
{Frebel}, A., \& {Norris}, J.~E. 2015, \araa, 53, 631

\bibitem[{{Frebel} {et~al.}(2005){Frebel}, {Aoki}, {Christlieb}, {Ando}, {Asplund}, {Barklem}, {Beers}, {Eriksson}, {Fechner}, {Fujimoto}, {Honda}, {Kajino}, {Minezaki}, {Nomoto}, {Norris}, {Ryan}, {Takada-Hidai}, {Tsangarides}, \& {Yoshii}}]{frebel2005}
{Frebel}, A., {Aoki}, W., {Christlieb}, N., {et~al.} 2005, \nat, 434, 871

\bibitem[{{Gaia Collaboration} {et~al.}(2016){Gaia Collaboration}, {Prusti}, {de Bruijne}, {Brown}, {Vallenari}, {Babusiaux}, {Bailer-Jones}, {Bastian}, {Biermann}, {Evans}, {Eyer}, {Jansen}, {Jordi}, {Klioner}, {Lammers}, {Lindegren}, {Luri}, {Mignard}, {Milligan}, {Panem}, {Poinsignon}, {Pourbaix}, {Randich}, {Sarri}, {Sartoretti}, {Siddiqui}, {Soubiran}, {Valette}, {van Leeuwen}, {Walton}, {Aerts}, {Arenou}, {Cropper}, {Drimmel}, {H{\o}g}, {Katz}, {Lattanzi}, {O'Mullane}, {Grebel}, {Holland}, {Huc}, {Passot}, {Bramante}, {Cacciari}, {Casta{\~n}eda}, {Chaoul}, {Cheek}, {De Angeli}, {Fabricius}, {Guerra}, {Hern{\'a}ndez}, {Jean-Antoine-Piccolo}, {Masana}, {Messineo}, {Mowlavi}, {Nienartowicz}, {Ord{\'o}{\~n}ez-Blanco}, {Panuzzo}, {Portell}, {Richards}, {Riello}, {Seabroke}, {Tanga}, {Th{\'e}venin}, {Torra}, {Els}, {Gracia-Abril}, {Comoretto}, {Garcia-Reinaldos}, {Lock}, {Mercier}, {Altmann}, {Andrae}, {Astraatmadja}, {Bellas-Velidis}, {Benson}, {Berthier}, {Blomme}, {Busso}, {Carry}, {Cellino}, {Clementini},
  {Cowell}, {Creevey}, {Cuypers}, {Davidson}, {De Ridder}, {de Torres}, {Delchambre}, {Dell'Oro}, {Ducourant}, {Fr{\'e}mat}, {Garc{\'\i}a-Torres}, {Gosset}, {Halbwachs}, {Hambly}, {Harrison}, {Hauser}, {Hestroffer}, {Hodgkin}, {Huckle}, {Hutton}, {Jasniewicz}, {Jordan}, {Kontizas}, {Korn}, {Lanzafame}, {Manteiga}, {Moitinho}, {Muinonen}, {Osinde}, {Pancino}, {Pauwels}, {Petit}, {Recio-Blanco}, {Robin}, {Sarro}, {Siopis}, {Smith}, {Smith}, {Sozzetti}, {Thuillot}, {van Reeven}, {Viala}, {Abbas}, {Abreu Aramburu}, {Accart}, {Aguado}, {Allan}, {Allasia}, {Altavilla}, {{\'A}lvarez}, {Alves}, {Anderson}, {Andrei}, {Anglada Varela}, {Antiche}, {Antoja}, {Ant{\'o}n}, {Arcay}, {Atzei}, {Ayache}, {Bach}, {Baker}, {Balaguer-N{\'u}{\~n}ez}, {Barache}, {Barata}, {Barbier}, {Barblan}, {Baroni}, {Barrado y Navascu{\'e}s}, {Barros}, {Barstow}, {Becciani}, {Bellazzini}, {Bellei}, {Bello Garc{\'\i}a}, {Belokurov}, {Bendjoya}, {Berihuete}, {Bianchi}, {Bienaym{\'e}}, {Billebaud}, {Blagorodnova}, {Blanco-Cuaresma}, {Boch},
  {Bombrun}, {Borrachero}, {Bouquillon}, {Bourda}, {Bouy}, {Bragaglia}, {Breddels}, {Brouillet}, {Br{\"u}semeister}, {Bucciarelli}, {Budnik}, {Burgess}, {Burgon}, {Burlacu}, {Busonero}, {Buzzi}, {Caffau}, {Cambras}, {Campbell}, {Cancelliere}, {Cantat-Gaudin}, {Carlucci}, {Carrasco}, {Castellani}, {Charlot}, {Charnas}, {Charvet}, {Chassat}, {Chiavassa}, {Clotet}, {Cocozza}, {Collins}, {Collins}, {Costigan}, {Crifo}, {Cross}, {Crosta}, {Crowley}, {Dafonte}, {Damerdji}, {Dapergolas}, {David}, {David}, {De Cat}, {de Felice}, {de Laverny}, {De Luise}, {De March}, {de Martino}, {de Souza}, {Debosscher}, {del Pozo}, {Delbo}, {Delgado}, {Delgado}, {di Marco}, {Di Matteo}, {Diakite}, {Distefano}, {Dolding}, {Dos Anjos}, {Drazinos}, {Dur{\'a}n}, {Dzigan}, {Ecale}, {Edvardsson}, {Enke}, {Erdmann}, {Escolar}, {Espina}, {Evans}, {Eynard Bontemps}, {Fabre}, {Fabrizio}, {Faigler}, {Falc{\~a}o}, {Farr{\`a}s Casas}, {Faye}, {Federici}, {Fedorets}, {Fern{\'a}ndez-Hern{\'a}ndez}, {Fernique}, {Fienga}, {Figueras}, {Filippi},
  {Findeisen}, {Fonti}, {Fouesneau}, {Fraile}, {Fraser}, {Fuchs}, {Furnell}, {Gai}, {Galleti}, {Galluccio}, {Garabato}, {Garc{\'\i}a-Sedano}, {Gar{\'e}}, {Garofalo}, {Garralda}, {Gavras}, {Gerssen}, {Geyer}, {Gilmore}, {Girona}, {Giuffrida}, {Gomes}, {Gonz{\'a}lez-Marcos}, {Gonz{\'a}lez-N{\'u}{\~n}ez}, {Gonz{\'a}lez-Vidal}, {Granvik}, {Guerrier}, {Guillout}, {Guiraud}, {G{\'u}rpide}, {Guti{\'e}rrez-S{\'a}nchez}, {Guy}, {Haigron}, {Hatzidimitriou}, {Haywood}, {Heiter}, {Helmi}, {Hobbs}, {Hofmann}, {Holl}, {Holland }, {Hunt}, {Hypki}, {Icardi}, {Irwin}, {Jevardat de Fombelle}, {Jofr{\'e}}, {Jonker}, {Jorissen}, {Julbe}, {Karampelas}, {Kochoska}, {Kohley}, {Kolenberg}, {Kontizas}, {Koposov}, {Kordopatis}, {Koubsky}, {Kowalczyk}, {Krone-Martins}, {Kudryashova}, {Kull}, {Bachchan}, {Lacoste-Seris}, {Lanza}, {Lavigne}, {Le Poncin-Lafitte}, {Lebreton}, {Lebzelter}, {Leccia}, {Leclerc}, {Lecoeur-Taibi}, {Lemaitre}, {Lenhardt}, {Leroux}, {Liao}, {Licata}, {Lindstr{\o}m}, {Lister}, {Livanou}, {Lobel}, {L{\"o}ffler},
  {L{\'o}pez}, {Lopez-Lozano}, {Lorenz}, {Loureiro}, {MacDonald}, {Magalh{\~a}es Fernandes}, {Managau}, {Mann}, {Mantelet}, {Marchal}, {Marchant}, {Marconi}, {Marie}, {Marinoni}, {Marrese}, {Marschalk{\'o}}, {Marshall}, {Mart{\'\i}n-Fleitas}, {Martino}, {Mary}, {Matijevi{\v{c}}}, {Mazeh}, {McMillan}, {Messina}, {Mestre}, {Michalik}, {Millar}, {Miranda}, {Molina}, {Molinaro}, {Molinaro}, {Moln{\'a}r}, {Moniez}, {Montegriffo}, {Monteiro}, {Mor}, {Mora}, {Morbidelli}, {Morel}, {Morgenthaler}, {Morley}, {Morris}, {Mulone}, {Muraveva}, {Musella}, {Narbonne}, {Nelemans}, {Nicastro}, {Noval}, {Ord{\'e}novic}, {Ordieres-Mer{\'e}}, {Osborne}, {Pagani}, {Pagano}, {Pailler}, {Palacin}, {Palaversa}, {Parsons}, {Paulsen}, {Pecoraro}, {Pedrosa}, {Pentik{\"a}inen}, {Pereira}, {Pichon}, {Piersimoni}, {Pineau}, {Plachy}, {Plum}, {Poujoulet}, {Pr{\v{s}}a}, {Pulone}, {Ragaini}, {Rago}, {Rambaux}, {Ramos-Lerate}, {Ranalli}, {Rauw}, {Read}, {Regibo}, {Renk}, {Reyl{\'e}}, {Ribeiro}, {Rimoldini}, {Ripepi}, {Riva}, {Rixon},
  {Roelens}, {Romero-G{\'o}mez}, {Rowell}, {Royer}, {Rudolph}, {Ruiz-Dern}, {Sadowski}, {Sagrist{\`a} Sell{\'e}s}, {Sahlmann}, {Salgado}, {Salguero}, {Sarasso}, {Savietto}, {Schnorhk}, {Schultheis}, {Sciacca}, {Segol}, {Segovia}, {Segransan}, {Serpell}, {Shih}, {Smareglia}, {Smart}, {Smith}, {Solano}, {Solitro}, {Sordo}, {Soria Nieto}, {Souchay}, {Spagna}, {Spoto}, {Stampa}, {Steele}, {Steidelm{\"u}ller}, {Stephenson}, {Stoev}, {Suess}, {S{\"u}veges}, {Surdej}, {Szabados}, {Szegedi-Elek}, {Tapiador}, {Taris}, {Tauran}, {Taylor}, {Teixeira}, {Terrett}, {Tingley}, {Trager}, {Turon}, {Ulla}, {Utrilla}, {Valentini}, {van Elteren}, {Van Hemelryck}, {van Leeuwen}, {Varadi}, {Vecchiato}, {Veljanoski}, {Via}, {Vicente}, {Vogt}, {Voss}, {Votruba}, {Voutsinas}, {Walmsley}, {Weiler}, {Weingrill}, {Werner}, {Wevers}, {Whitehead}, {Wyrzykowski}, {Yoldas}, {{\v{Z}}erjal}, {Zucker}, {Zurbach}, {Zwitter}, {Alecu}, {Allen}, {Allende Prieto}, {Amorim}, {Anglada-Escud{\'e}}, {Arsenijevic}, {Azaz}, {Balm}, {Beck}, {Bernstein},
  {Bigot}, {Bijaoui}, {Blasco}, {Bonfigli}, {Bono}, {Boudreault}, {Bressan}, {Brown}, {Brunet}, {Bunclark}, {Buonanno}, {Butkevich}, {Carret}, {Carrion}, {Chemin}, {Ch{\'e}reau}, {Corcione}, {Darmigny}, {de Boer}, {de Teodoro}, {de Zeeuw}, {Delle Luche}, {Domingues}, {Dubath}, {Fodor}, {Fr{\'e}zouls}, {Fries}, {Fustes}, {Fyfe}, {Gallardo}, {Gallegos}, {Gardiol}, {Gebran}, {Gomboc}, {G{\'o}mez}, {Grux}, {Gueguen}, {Heyrovsky}, {Hoar}, {Iannicola}, {Isasi Parache}, {Janotto}, {Joliet}, {Jonckheere}, {Keil}, {Kim}, {Klagyivik}, {Klar}, {Knude}, {Kochukhov}, {Kolka}, {Kos}, {Kutka}, {Lainey}, {LeBouquin}, {Liu}, {Loreggia}, {Makarov}, {Marseille}, {Martayan}, {Martinez-Rubi}, {Massart}, {Meynadier}, {Mignot}, {Munari}, {Nguyen}, {Nordlander}, {Ocvirk}, {O'Flaherty}, {Olias Sanz}, {Ortiz}, {Osorio}, {Oszkiewicz}, {Ouzounis}, {Palmer}, {Park}, {Pasquato}, {Peltzer}, {Peralta}, {P{\'e}turaud}, {Pieniluoma}, {Pigozzi}, {Poels}, {Prat}, {Prod'homme}, {Raison}, {Rebordao}, {Risquez}, {Rocca-Volmerange}, {Rosen},
  {Ruiz-Fuertes}, {Russo}, {Sembay}, {Serraller Vizcaino}, {Short}, {Siebert}, {Silva}, {Sinachopoulos}, {Slezak}, {Soffel}, {Sosnowska}, {Strai{\v{z}}ys}, {ter Linden}, {Terrell}, {Theil}, {Tiede}, {Troisi}, {Tsalmantza}, {Tur}, {Vaccari}, {Vachier}, {Valles}, {Van Hamme}, {Veltz}, {Virtanen}, {Wallut}, {Wichmann}, {Wilkinson}, {Ziaeepour}, \& {Zschocke}}]{GaiaMission}
{Gaia Collaboration}, {Prusti}, T., {de Bruijne}, J.~H.~J., {et~al.} 2016, \aap, 595, A1

\bibitem[{{Gaia Collaboration} {et~al.}(2023){Gaia Collaboration}, {Vallenari}, {Brown}, {Prusti}, {de Bruijne}, {Arenou}, {Babusiaux}, {Biermann}, {Creevey}, {Ducourant}, {Evans}, {Eyer}, {Guerra}, {Hutton}, {Jordi}, {Klioner}, {Lammers}, {Lindegren}, {Luri}, {Mignard}, {Panem}, {Pourbaix}, {Randich}, {Sartoretti}, {Soubiran}, {Tanga}, {Walton}, {Bailer-Jones}, {Bastian}, {Drimmel}, {Jansen}, {Katz}, {Lattanzi}, {van Leeuwen}, {Bakker}, {Cacciari}, {Casta{\~n}eda}, {De Angeli}, {Fabricius}, {Fouesneau}, {Fr{\'e}mat}, {Galluccio}, {Guerrier}, {Heiter}, {Masana}, {Messineo}, {Mowlavi}, {Nicolas}, {Nienartowicz}, {Pailler}, {Panuzzo}, {Riclet}, {Roux}, {Seabroke}, {Sordo}, {Th{\'e}venin}, {Gracia-Abril}, {Portell}, {Teyssier}, {Altmann}, {Andrae}, {Audard}, {Bellas-Velidis}, {Benson}, {Berthier}, {Blomme}, {Burgess}, {Busonero}, {Busso}, {C{\'a}novas}, {Carry}, {Cellino}, {Cheek}, {Clementini}, {Damerdji}, {Davidson}, {de Teodoro}, {Nu{\~n}ez Campos}, {Delchambre}, {Dell'Oro}, {Esquej},
  {Fern{\'a}ndez-Hern{\'a}ndez}, {Fraile}, {Garabato}, {Garc{\'\i}a-Lario}, {Gosset}, {Haigron}, {Halbwachs}, {Hambly}, {Harrison}, {Hern{\'a}ndez}, {Hestroffer}, {Hodgkin}, {Holl}, {Jan{\ss}en}, {Jevardat de Fombelle}, {Jordan}, {Krone-Martins}, {Lanzafame}, {L{\"o}ffler}, {Marchal}, {Marrese}, {Moitinho}, {Muinonen}, {Osborne}, {Pancino}, {Pauwels}, {Recio-Blanco}, {Reyl{\'e}}, {Riello}, {Rimoldini}, {Roegiers}, {Rybizki}, {Sarro}, {Siopis}, {Smith}, {Sozzetti}, {Utrilla}, {van Leeuwen}, {Abbas}, {{\'A}brah{\'a}m}, {Abreu Aramburu}, {Aerts}, {Aguado}, {Ajaj}, {Aldea-Montero}, {Altavilla}, {{\'A}lvarez}, {Alves}, {Anders}, {Anderson}, {Anglada Varela}, {Antoja}, {Baines}, {Baker}, {Balaguer-N{\'u}{\~n}ez}, {Balbinot}, {Balog}, {Barache}, {Barbato}, {Barros}, {Barstow}, {Bartolom{\'e}}, {Bassilana}, {Bauchet}, {Becciani}, {Bellazzini}, {Berihuete}, {Bernet}, {Bertone}, {Bianchi}, {Binnenfeld}, {Blanco-Cuaresma}, {Blazere}, {Boch}, {Bombrun}, {Bossini}, {Bouquillon}, {Bragaglia}, {Bramante}, {Breedt},
  {Bressan}, {Brouillet}, {Brugaletta}, {Bucciarelli}, {Burlacu}, {Butkevich}, {Buzzi}, {Caffau}, {Cancelliere}, {Cantat-Gaudin}, {Carballo}, {Carlucci}, {Carnerero}, {Carrasco}, {Casamiquela}, {Castellani}, {Castro-Ginard}, {Chaoul}, {Charlot}, {Chemin}, {Chiaramida}, {Chiavassa}, {Chornay}, {Comoretto}, {Contursi}, {Cooper}, {Cornez}, {Cowell}, {Crifo}, {Cropper}, {Crosta}, {Crowley}, {Dafonte}, {Dapergolas}, {David}, {David}, {de Laverny}, {De Luise}, \& {De March}}]{GaiaDR3}
{Gaia Collaboration}, {Vallenari}, A., {Brown}, A.~G.~A., {et~al.} 2023, \aap, 674, A1

\bibitem[{{Galarza} {et~al.}(2022){Galarza}, {Daflon}, {Placco}, {Allende Prieto}, {Borges Fernandes}, {Yuan}, {L{\'o}pez-Sanjuan}, {Lee}, {Solano}, {Jim{\'e}nez-Esteban}, {Sobral}, {Alvarez Candal}, {Pereira}, {Akras}, {Mart{\'\i}n}, {Jim{\'e}nez Teja}, {Cenarro}, {Crist{\'o}bal-Hornillos}, {Hern{\'a}ndez-Monteagudo}, {Mar{\'\i}n-Franch}, {Moles}, {Varela}, {V{\'a}zquez Rami{\'o}}, {Alcaniz}, {Dupke}, {Ederoclite}, {Sodr{\'e}}, \& {Angulo}}]{Galarza2022jplus}
{Galarza}, C.~A., {Daflon}, S., {Placco}, V.~M., {et~al.} 2022, \aap, 657, A35

\bibitem[{{Gimeno} {et~al.}(2016){Gimeno}, {Roth}, {Chiboucas}, {Hibon}, {Boucher}, {White}, {Rippa}, {Labrie}, {Turner}, {Hanna}, {Lazo}, {P{\'e}rez}, {Rogers}, {Rojas}, {Placco}, \& {Murowinski}}]{GMOSPaper2}
{Gimeno}, G., {Roth}, K., {Chiboucas}, K., {et~al.} 2016, in Society of Photo-Optical Instrumentation Engineers (SPIE) Conference Series, Vol. 9908, Ground-based and Airborne Instrumentation for Astronomy VI, ed. C.~J. {Evans}, L.~{Simard}, \& H.~{Takami}, 99082S

\bibitem[{{Gravity Collaboration} {et~al.}(2018){Gravity Collaboration}, {Abuter}, {Amorim}, {Anugu}, {Baub{\"o}ck}, {Benisty}, {Berger}, {Blind}, {Bonnet}, {Brandner}, {Buron}, {Collin}, {Chapron}, {Cl{\'e}net}, {Coud{\'e} Du Foresto}, {de Zeeuw}, {Deen}, {Delplancke-Str{\"o}bele}, {Dembet}, {Dexter}, {Duvert}, {Eckart}, {Eisenhauer}, {Finger}, {F{\"o}rster Schreiber}, {F{\'e}dou}, {Garcia}, {Garcia Lopez}, {Gao}, {Gendron}, {Genzel}, {Gillessen}, {Gordo}, {Habibi}, {Haubois}, {Haug}, {Hau{\ss}mann}, {Henning}, {Hippler}, {Horrobin}, {Hubert}, {Hubin}, {Jimenez Rosales}, {Jochum}, {Jocou}, {Kaufer}, {Kellner}, {Kendrew}, {Kervella}, {Kok}, {Kulas}, {Lacour}, {Lapeyr{\`e}re}, {Lazareff}, {Le Bouquin}, {L{\'e}na}, {Lippa}, {Lenzen}, {M{\'e}rand}, {M{\"u}ler}, {Neumann}, {Ott}, {Palanca}, {Paumard}, {Pasquini}, {Perraut}, {Perrin}, {Pfuhl}, {Plewa}, {Rabien}, {Ram{\'\i}rez}, {Ramos}, {Rau}, {Rodr{\'\i}guez-Coira}, {Rohloff}, {Rousset}, {Sanchez-Bermudez}, {Scheithauer}, {Sch{\"o}ller}, {Schuler}, {Spyromilio},
  {Straub}, {Straubmeier}, {Sturm}, {Tacconi}, {Tristram}, {Vincent}, {von Fellenberg}, {Wank}, {Waisberg}, {Widmann}, {Wieprecht}, {Wiest}, {Wiezorrek}, {Woillez}, {Yazici}, {Ziegler}, \& {Zins}}]{GRAVITY2018}
{Gravity Collaboration}, {Abuter}, R., {Amorim}, A., {et~al.} 2018, \aap, 615, L15

\bibitem[{{Hansen} {et~al.}(2014){Hansen}, {Hansen}, {Christlieb}, {Yong}, {Bessell}, {Garc{\'\i}a P{\'e}rez}, {Beers}, {Placco}, {Frebel}, {Norris}, \& {Asplund}}]{hansen2014}
{Hansen}, T., {Hansen}, C.~J., {Christlieb}, N., {et~al.} 2014, \apj, 787, 162

\bibitem[{{Heger} \& {Woosley}(2010)}]{Heger2010}
{Heger}, A., \& {Woosley}, S.~E. 2010, \apj, 724, 341

\bibitem[{{Henden} {et~al.}(2016){Henden}, {Templeton}, {Terrell}, {Smith}, {Levine}, \& {Welch}}]{apassdr9}
{Henden}, A.~A., {Templeton}, M., {Terrell}, D., {et~al.} 2016, VizieR Online Data Catalog, II/336

\bibitem[{{Holmbeck} {et~al.}(2020){Holmbeck}, {Hansen}, {Beers}, {Placco}, {Whitten}, {Rasmussen}, {Roederer}, {Ezzeddine}, {Sakari}, {Frebel}, {Drout}, {Simon}, {Thompson}, {Bland-Hawthorn}, {Gibson}, {Grebel}, {Kordopatis}, {Kunder}, {Mel{\'e}ndez}, {Navarro}, {Reid}, {Seabroke}, {Steinmetz}, {Watson}, \& {Wyse}}]{holmbeck2020}
{Holmbeck}, E.~M., {Hansen}, T.~T., {Beers}, T.~C., {et~al.} 2020, \apjs, 249, 30

\bibitem[{{Ishigaki} {et~al.}(2018){Ishigaki}, {Tominaga}, {Kobayashi}, \& {Nomoto}}]{Ishigaki2018}
{Ishigaki}, M.~N., {Tominaga}, N., {Kobayashi}, C., \& {Nomoto}, K. 2018, \apj, 857, 46

\bibitem[{{Jacobson} {et~al.}(2015){Jacobson}, {Keller}, {Frebel}, {Casey}, {Asplund}, {Bessell}, {Da Costa}, {Lind}, {Marino}, {Norris}, {Pe{\~n}a}, {Schmidt}, {Tisserand}, {Walsh}, {Yong}, \& {Yu}}]{Jacobson2015metalpoor}
{Jacobson}, H.~R., {Keller}, S., {Frebel}, A., {et~al.} 2015, \apj, 807, 171

\bibitem[{{Ji} {et~al.}(2020{\natexlab{a}}){Ji}, {Li}, {Simon}, {Marshall}, {Vivas}, {Pace}, {Bechtol}, {Drlica-Wagner}, {Koposov}, {Hansen}, {Allam}, {Gruendl}, {Johnson}, {McNanna}, {No{\"e}l}, {Tucker}, \& {Walker}}]{Ji2020MagLiteS}
{Ji}, A.~P., {Li}, T.~S., {Simon}, J.~D., {et~al.} 2020{\natexlab{a}}, \apj, 889, 27

\bibitem[{{Ji} {et~al.}(2020{\natexlab{b}}){Ji}, {Li}, {Hansen}, {Casey}, {Koposov}, {Pace}, {Mackey}, {Lewis}, {Simpson}, {Bland-Hawthorn}, {Cullinane}, {Da Costa}, {Hattori}, {Martell}, {Kuehn}, {Erkal}, {Shipp}, {Wan}, \& {Zucker}}]{Ji2020streams}
{Ji}, A.~P., {Li}, T.~S., {Hansen}, T.~T., {et~al.} 2020{\natexlab{b}}, \aj, 160, 181

\bibitem[{{Ji} {et~al.}(2023){Ji}, {Simon}, {Roederer}, {Magg}, {Frebel}, {Johnson}, {Klessen}, {Magg}, {Cescutti}, {Mateo}, {Bergemann}, \& {Bailey}}]{AlexJi2023ret2}
{Ji}, A.~P., {Simon}, J.~D., {Roederer}, I.~U., {et~al.} 2023, \aj, 165, 100

\bibitem[{{Johnson} {et~al.}(2020){Johnson}, {Conroy}, {Naidu}, {Bonaca}, {Zaritsky}, {Ting}, {Cargile}, {Han}, \& {Speagle}}]{Johnson2020sgr}
{Johnson}, B.~D., {Conroy}, C., {Naidu}, R.~P., {et~al.} 2020, \apj, 900, 103

\bibitem[{{Karovicova} {et~al.}(2020){Karovicova}, {White}, {Nordlander}, {Casagrande}, {Ireland}, {Huber}, \& {Jofr{\'e}}}]{Karovicova2020standards}
{Karovicova}, I., {White}, T.~R., {Nordlander}, T., {et~al.} 2020, \aap, 640, A25

\bibitem[{{Keller} {et~al.}(2014){Keller}, {Bessell}, {Frebel}, {Casey}, {Asplund}, {Jacobson}, {Lind}, {Norris}, {Yong}, {Heger}, {Magic}, {da Costa}, {Schmidt}, \& {Tisserand}}]{keller2014}
{Keller}, S.~C., {Bessell}, M.~S., {Frebel}, A., {et~al.} 2014, \nat, 506, 463

\bibitem[{{Kelson}(2003)}]{carpy}
{Kelson}, D.~D. 2003, \pasp, 115, 688

\bibitem[{{Labrie} {et~al.}(2023){Labrie}, {Simpson}, {Cardenes}, {Turner}, {Soraisam}, {Quint}, {Oberdorf}, {Placco}, {Berke}, {Smirnova}, {Conseil}, {Vacca}, \& {Thomas-Osip}}]{DRAGONS2}
{Labrie}, K., {Simpson}, C., {Cardenes}, R., {et~al.} 2023, Research Notes of the American Astronomical Society, 7, 214

\bibitem[{{Li} {et~al.}(2022){Li}, {Aoki}, {Matsuno}, {Xing}, {Suda}, {Tominaga}, {Chen}, {Honda}, {Ishigaki}, {Shi}, {Zhao}, \& {Zhao}}]{Li2022lamost}
{Li}, H., {Aoki}, W., {Matsuno}, T., {et~al.} 2022, \apj, 931, 147

\bibitem[{{Limberg} {et~al.}(2023){Limberg}, {Queiroz}, {Perottoni}, {Rossi}, {Amarante}, {Santucci}, {Chiappini}, {P{\'e}rez-Villegas}, \& {Lee}}]{Limberg2023sgr}
{Limberg}, G., {Queiroz}, A. B.~A., {Perottoni}, H.~D., {et~al.} 2023, \apj, 946, 66

\bibitem[{{Mardini} {et~al.}(2024){Mardini}, {Frebel}, \& {Chiti}}]{mardini2024diskUMP}
{Mardini}, M.~K., {Frebel}, A., \& {Chiti}, A. 2024, \mnras, 529, L60

\bibitem[{{Mardini} {et~al.}(2022){Mardini}, {Frebel}, {Chiti}, {Meiron}, {Brauer}, \& {Ou}}]{mardini2022atari}
{Mardini}, M.~K., {Frebel}, A., {Chiti}, A., {et~al.} 2022, \apj, 936, 78

\bibitem[{{McMillan}(2017)}]{mcmillan2017}
{McMillan}, P.~J. 2017, \mnras, 465, 76

\bibitem[{{Mendes de Oliveira} {et~al.}(2019){Mendes de Oliveira}, {Ribeiro}, {Schoenell}, {Kanaan}, {Overzier}, {Molino}, {Sampedro}, {Coelho}, {Barbosa}, {Cortesi}, {Costa-Duarte}, {Herpich}, {Hernand ez-Jimenez}, {Placco}, {Xavier}, {Abramo}, {Saito}, {Chies-Santos}, {Ederoclite}, {Lopes de Oliveira}, {Gon{\c{c}}alves}, {Akras}, {Almeida}, {Almeida-Fernandes}, {Beers}, {Bonatto}, {Bonoli}, {Cypriano}, {Vinicius-Lima}, {de Souza}, {Fabiano de Souza}, {Ferrari}, {Gon{\c{c}}alves}, {Gonzalez}, {Guti{\'e}rrez-Soto}, {Hartmann}, {Jaffe}, {Kerber}, {Lima-Dias}, {Lopes}, {Menendez-Delmestre}, {Nakazono}, {Novais}, {Ortega-Minakata}, {Pereira}, {Perottoni}, {Queiroz}, {Reis}, {Santos}, {Santos-Silva}, {Santucci}, {Barbosa}, {Siffert}, {Sodr{\'e}}, {Torres-Flores}, {Westera}, {Whitten}, {Alcaniz}, {Alonso-Garc{\'\i}a}, {Alencar}, {Alvarez-Cand al}, {Amram}, {Azanha}, {Barb{\'a}}, {Bernardinelli}, {Borges Fernandes}, {Branco}, {Brito-Silva}, {Buzzo}, {Caffer}, {Campillay}, {Cano}, {Carvano}, {Castejon}, {Cid
  Fernandes}, {Dantas}, {Daflon}, {Damke}, {de la Reza}, {de Melo de Azevedo}, {De Paula}, {Diem}, {Donnerstein}, {Dors}, {Dupke}, {Eikenberry}, {Escudero}, {Faifer}, {Far{\'\i}as}, {Fernandes}, {Fernandes}, {Fontes}, {Galarza}, {Hirata}, {Katena}, {Gregorio-Hetem}, {Hern{\'a}ndez-Fern{\'a}ndez}, {Izzo}, {Jaque Arancibia}, {Jatenco-Pereira}, {Jim{\'e}nez-Teja}, {Kann}, {Krabbe}, {Labayru}, {Lazzaro}, {Lima Neto}, {Lopes}, {Magalh{\~a}es}, {Makler}, {de Menezes}, {Miralda-Escud{\'e}}, {Monteiro-Oliveira}, {Montero-Dorta}, {Mu{\~n}oz-Elgueta}, {Nemmen}, {Nilo Castell{\'o}n}, {Oliveira}, {Ort{\'\i}z}, {Pattaro}, {Pereira}, {Quint}, {Riguccini}, {Rocha Pinto}, {Rodrigues}, {Roig}, {Rossi}, {Saha}, {Santos}, {Schnorr M{\"u}ller}, {Sesto}, {Silva}, {Smith Castelli}, {Teixeira}, {Telles}, {Thom de Souza}, {Th{\"o}ne}, {Trevisan}, {de Ugarte Postigo}, {Urrutia-Viscarra}, {Veiga}, {Vika}, {Vitorelli}, {Werle}, {Werner}, \& {Zaritsky}}]{splus}
{Mendes de Oliveira}, C., {Ribeiro}, T., {Schoenell}, W., {et~al.} 2019, MNRAS, 489, 241

\bibitem[{{Mucciarelli} {et~al.}(2021){Mucciarelli}, {Bellazzini}, \& {Massari}}]{mucciarelli2021calib}
{Mucciarelli}, A., {Bellazzini}, M., \& {Massari}, D. 2021, \aap, 653, A90

\bibitem[{{Nidever} {et~al.}(2010){Nidever}, {Majewski}, {Butler Burton}, \& {Nigra}}]{Nidever2010magstream}
{Nidever}, D.~L., {Majewski}, S.~R., {Butler Burton}, W., \& {Nigra}, L. 2010, \apj, 723, 1618

\bibitem[{{Nordlander} {et~al.}(2019){Nordlander}, {Bessell}, {Da Costa}, {Mackey}, {Asplund}, {Casey}, {Chiti}, {Ezzeddine}, {Frebel}, {Lind}, {Marino}, {Murphy}, {Norris}, {Schmidt}, \& {Yong}}]{nordlander2019smss}
{Nordlander}, T., {Bessell}, M.~S., {Da Costa}, G.~S., {et~al.} 2019, \mnras, 488, L109

\bibitem[{{Norris} \& {Yong}(2019)}]{NorrisYong2019nlte}
{Norris}, J.~E., \& {Yong}, D. 2019, \apj, 879, 37

\bibitem[{{Onken} {et~al.}(2024){Onken}, {Wolf}, {Bessell}, {Chang}, {Luvaul}, {Tonry}, {White}, \& {Da Costa}}]{SkyMapperDR4}
{Onken}, C.~A., {Wolf}, C., {Bessell}, M.~S., {et~al.} 2024, \pasa, 41, e061

\bibitem[{{Pace}(2024)}]{pace2024lvdb}
{Pace}, A.~B. 2024, arXiv e-prints, arXiv:2411.07424

\bibitem[{{Perottoni} {et~al.}(2024){Perottoni}, {Placco}, {Almeida-Fernandes}, {Herpich}, {Rossi}, {Beers}, {Smiljanic}, {Amarante}, {Limberg}, {Werle}, {Rocha-Pinto}, {Beraldo e Silva}, {Daflon}, {Alvarez-Candal}, {Oliveira Schwarz}, {Schoenell}, {Ribeiro}, \& {Kanaan}}]{splusSHORTSdr1}
{Perottoni}, H.~D., {Placco}, V.~M., {Almeida-Fernandes}, F., {et~al.} 2024, \aap, 691, A138

\bibitem[{{Placco} {et~al.}(2022){Placco}, {Almeida-Fernandes}, {Arentsen}, {Lee}, {Schoenell}, {Ribeiro}, \& {Kanaan}}]{placco2022splus}
{Placco}, V.~M., {Almeida-Fernandes}, F., {Arentsen}, A., {et~al.} 2022, \apjs, 262, 8

\bibitem[{{Placco} {et~al.}(2014){Placco}, {Frebel}, {Beers}, \& {Stancliffe}}]{placco2014Carbon}
{Placco}, V.~M., {Frebel}, A., {Beers}, T.~C., \& {Stancliffe}, R.~J. 2014, \apj, 797, 21

\bibitem[{{Placco} {et~al.}(2021{\natexlab{a}}){Placco}, {Sneden}, {Roederer}, {Lawler}, {Den Hartog}, {Hejazi}, {Maas}, \& {Bernath}}]{Placco2021linemake}
{Placco}, V.~M., {Sneden}, C., {Roederer}, I.~U., {et~al.} 2021{\natexlab{a}}, Research Notes of the American Astronomical Society, 5, 92

\bibitem[{{Placco} {et~al.}(2021{\natexlab{b}}){Placco}, {Roederer}, {Lee}, {Almeida-Fernandes}, {Herpich}, {Perottoni}, {Schoenell}, {Ribeiro}, \& {Kanaan}}]{placco2021ump}
{Placco}, V.~M., {Roederer}, I.~U., {Lee}, Y.~S., {et~al.} 2021{\natexlab{b}}, \apjl, 912, L32

\bibitem[{{Placco} {et~al.}(2024){Placco}, {Gupta}, {Almeida-Fernandes}, {Logsdon}, {Rajagopal}, {Holmbeck}, {Roederer}, {Della Costa}, {Fernandez}, {Golub}, {Higuera}, {Patel}, {Ridgway}, \& {Schweiker}}]{placco2024}
{Placco}, V.~M., {Gupta}, A.~F., {Almeida-Fernandes}, F., {et~al.} 2024, \apj, 977, 12

\bibitem[{{Placco} {et~al.}(2025){Placco}, {Limberg}, {Chiti}, {Prabhu}, {Ji}, {Barbosa}, {Cerny}, {Pace}, {Stringfellow}, {Sand}, {Mart{\'\i}nez-V{\'a}zquez}, {Riley}, {Rossi}, {No{\"e}l}, {Vivas}, {Medina}, {Drlica-Wagner}, {Sakowska}, {Mutlu-Pakdil}, {Massana}, {Carballo-Bello}, {Choi}, {Crnojevi{\'c}}, \& {Tan}}]{placco2025magic}
{Placco}, V.~M., {Limberg}, G., {Chiti}, A., {et~al.} 2025, arXiv e-prints, arXiv:2506.19163

\bibitem[{{Popa} {et~al.}(2023){Popa}, {Hoppe}, {Bergemann}, {Hansen}, {Plez}, \& {Beers}}]{popa2023nlteCH}
{Popa}, S.~A., {Hoppe}, R., {Bergemann}, M., {et~al.} 2023, \aap, 670, A25

\bibitem[{{Queiroz} {et~al.}(2018){Queiroz}, {Anders}, {Santiago}, {Chiappini}, {Steinmetz}, {Dal Ponte}, {Stassun}, {da Costa}, {Maia}, {Crestani}, {Beers}, {Fern{\'a}ndez-Trincado}, {Garc{\'\i}a-Hern{\'a}ndez}, {Roman-Lopes}, \& {Zamora}}]{Queiroz2018}
{Queiroz}, A.~B.~A., {Anders}, F., {Santiago}, B.~X., {et~al.} 2018, \mnras, 476, 2556

\bibitem[{{Roederer} {et~al.}(2014){Roederer}, {Preston}, {Thompson}, {Shectman}, {Sneden}, {Burley}, \& {Kelson}}]{roederer2014}
{Roederer}, I.~U., {Preston}, G.~W., {Thompson}, I.~B., {et~al.} 2014, \aj, 147, 136

\bibitem[{{Roederer} {et~al.}(2018){Roederer}, {Sakari}, {Placco}, {Beers}, {Ezzeddine}, {Frebel}, \& {Hansen}}]{roederer2018hd222925}
{Roederer}, I.~U., {Sakari}, C.~M., {Placco}, V.~M., {et~al.} 2018, \apj, 865, 129

\bibitem[{{Roman}(1950)}]{Roman1950}
{Roman}, N.~G. 1950, \apj, 112, 554

\bibitem[{{Rossi} {et~al.}(1999){Rossi}, {Beers}, \& {Sneden}}]{rossi1999}
{Rossi}, S., {Beers}, T.~C., \& {Sneden}, C. 1999, in Astronomical Society of the Pacific Conference Series, Vol. 165, The Third Stromlo Symposium: The Galactic Halo, ed. B.~K. {Gibson}, R.~S. {Axelrod}, \& M.~E. {Putman}, 264

\bibitem[{{Santiago} {et~al.}(2016){Santiago}, {Brauer}, {Anders}, {Chiappini}, {Queiroz}, {Girardi}, {Rocha-Pinto}, {Balbinot}, {da Costa}, {Maia}, {Schultheis}, {Steinmetz}, {Miglio}, {Montalb{\'a}n}, {Schneider}, {Beers}, {Frinchaboy}, {Lee}, \& {Zasowski}}]{Santiago2016starhorse}
{Santiago}, B.~X., {Brauer}, D.~E., {Anders}, F., {et~al.} 2016, \aap, 585, A42

\bibitem[{{Schlafly} \& {Finkbeiner}(2011)}]{Schlafly2011}
{Schlafly}, E.~F., \& {Finkbeiner}, D.~P. 2011, \apj, 737, 103

\bibitem[{{Schlegel} {et~al.}(1998){Schlegel}, {Finkbeiner}, \& {Davis}}]{Schlegel1998}
{Schlegel}, D.~J., {Finkbeiner}, D.~P., \& {Davis}, M. 1998, \apj, 500, 525

\bibitem[{{Skrutskie} {et~al.}(2006){Skrutskie}, {Cutri}, {Stiening}, {Weinberg}, {Schneider}, {Carpenter}, {Beichman}, {Capps}, {Chester}, {Elias}, {Huchra}, {Liebert}, {Lonsdale}, {Monet}, {Price}, {Seitzer}, {Jarrett}, {Kirkpatrick}, {Gizis}, {Howard}, {Evans}, {Fowler}, {Fullmer}, {Hurt}, {Light}, {Kopan}, {Marsh}, {McCallon}, {Tam}, {Van Dyk}, \& {Wheelock}}]{2MASS}
{Skrutskie}, M.~F., {Cutri}, R.~M., {Stiening}, R., {et~al.} 2006, \aj, 131, 1163

\bibitem[{{Sk{\'u}lad{\'o}ttir} {et~al.}(2021){Sk{\'u}lad{\'o}ttir}, {Salvadori}, {Amarsi}, {Tolstoy}, {Irwin}, {Hill}, {Jablonka}, {Battaglia}, {Starkenburg}, {Massari}, {Helmi}, \& {Posti}}]{Skuladottir2021UMPsculptor}
{Sk{\'u}lad{\'o}ttir}, {\'A}., {Salvadori}, S., {Amarsi}, A.~M., {et~al.} 2021, \apjl, 915, L30

\bibitem[{{Sneden}(1973)}]{Sneden1973moog}
{Sneden}, C.~A. 1973, PhD thesis, University of Texas, Austin

\bibitem[{{Sobeck} {et~al.}(2011){Sobeck}, {Kraft}, {Sneden}, {Preston}, {Cowan}, {Smith}, {Thompson}, {Shectman}, \& {Burley}}]{Sobeck2011}
{Sobeck}, J.~S., {Kraft}, R.~P., {Sneden}, C., {et~al.} 2011, \aj, 141, 175

\bibitem[{{Starkenburg} {et~al.}(2017){Starkenburg}, {Martin}, {Youakim}, {Aguado}, {Allende Prieto}, {Arentsen}, {Bernard}, {Bonifacio}, {Caffau}, {Carlberg}, {C{\^o}t{\'e}}, {Fouesneau}, {Fran{\c{c}}ois}, {Franke}, {Gonz{\'a}lez Hern{\'a}ndez}, {Gwyn}, {Hill}, {Ibata}, {Jablonka}, {Longeard}, {McConnachie}, {Navarro}, {S{\'a}nchez-Janssen}, {Tolstoy}, \& {Venn}}]{Starkenburg2017PRISTINE}
{Starkenburg}, E., {Martin}, N., {Youakim}, K., {et~al.} 2017, \mnras, 471, 2587

\bibitem[{{Starkenburg} {et~al.}(2018){Starkenburg}, {Aguado}, {Bonifacio}, {Caffau}, {Jablonka}, {Lardo}, {Martin}, {S{\'a}nchez-Janssen}, {Sestito}, {Venn}, {Youakim}, {Allende Prieto}, {Arentsen}, {Gentile}, {Gonz{\'a}lez Hern{\'a}ndez}, {Kielty}, {Koppelman}, {Longeard}, {Tolstoy}, {Carlberg}, {C{\^o}t{\'e}}, {Fouesneau}, {Hill}, {McConnachie}, \& {Navarro}}]{starkenburg2018ump}
{Starkenburg}, E., {Aguado}, D.~S., {Bonifacio}, P., {et~al.} 2018, \mnras, 481, 3838

\bibitem[{{Suda} {et~al.}(2008{\natexlab{a}}){Suda}, {Katsuta}, {Yamada}, {Suwa}, {Ishizuka}, {Komiya}, {Sorai}, {Aikawa}, \& {Fujimoto}}]{SAGA_1}
{Suda}, T., {Katsuta}, Y., {Yamada}, S., {et~al.} 2008{\natexlab{a}}, \pasj, 60, 1159

\bibitem[{{Suda} {et~al.}(2008{\natexlab{b}}){Suda}, {Katsuta}, {Yamada}, {Suwa}, {Ishizuka}, {Komiya}, {Sorai}, {Aikawa}, \& {Fujimoto}}]{suda2008}
---. 2008{\natexlab{b}}, \pasj, 60, 1159

\bibitem[{{Vasiliev}(2023)}]{vasiliev2023lmc}
{Vasiliev}, E. 2023, Galaxies, 11, 59

\bibitem[{{Vasiliev} {et~al.}(2021){Vasiliev}, {Belokurov}, \& {Erkal}}]{Vasiliev2021tango}
{Vasiliev}, E., {Belokurov}, V., \& {Erkal}, D. 2021, \mnras, 501, 2279

\bibitem[{{Venn} {et~al.}(2017){Venn}, {Starkenburg}, {Malo}, {Martin}, \& {Laevens}}]{venn2017graces}
{Venn}, K.~A., {Starkenburg}, E., {Malo}, L., {Martin}, N., \& {Laevens}, B.~P.~M. 2017, \mnras, 466, 3741

\bibitem[{{Wang} \& {Chen}(2019)}]{WangChen2019extinctions}
{Wang}, S., \& {Chen}, X. 2019, \apj, 877, 116

\bibitem[{{Whitten} {et~al.}(2019){Whitten}, {Placco}, {Beers}, {Chies-Santos}, {Bonatto}, {Varela}, {Crist{\'o}bal-Hornillos}, {Ederoclite}, {Masseron}, {Lee}, {Akras}, {Borges Fernandes}, {Caballero}, {Cenarro}, {Coelho}, {Costa-Duarte}, {Daflon}, {Dupke}, {Lopes de Oliveira}, {L{\'o}pez-Sanjuan}, {Mar{\'\i}n-Franch}, {Mendes de Oliveira}, {Moles}, {Orsi}, {Rossi}, {Sodr{\'e}}, \& {V{\'a}zquez Rami{\'o}}}]{Whitten2019jplus}
{Whitten}, D.~D., {Placco}, V.~M., {Beers}, T.~C., {et~al.} 2019, A\&A, 622, A182

\bibitem[{{Whitten} {et~al.}(2021){Whitten}, {Placco}, {Beers}, {An}, {Lee}, {Almeida-Fernandes}, {Herpich}, {Daflon}, {Barbosa}, {Perottoni}, {Rossi}, {Tissera}, {Yoon}, {Youakim}, {Schoenell}, {Ribeiro}, \& {Kanaan}}]{Whitten2021splus}
---. 2021, \apj, 912, 147

\bibitem[{{Yao} {et~al.}(2023){Yao}, {Ji}, {Koposov}, \& {Limberg}}]{Yao2023xp}
{Yao}, Y., {Ji}, A.~P., {Koposov}, S.~E., \& {Limberg}, G. 2023, arXiv e-prints, arXiv:2303.17676

\bibitem[{{Yi} {et~al.}(2001){Yi}, {Demarque}, {Kim}, {Lee}, {Ree}, {Lejeune}, \& {Barnes}}]{YaleYonsei2001}
{Yi}, S., {Demarque}, P., {Kim}, Y.-C., {et~al.} 2001, \apjs, 136, 417

\end{thebibliography}

\end{document}